\documentclass{article}

\usepackage{PRIMEarxiv}

\usepackage[T1]{fontenc}    
\usepackage{nicefrac}       
\usepackage{fancyhdr}       

\RequirePackage{amsthm,amsmath,amsfonts,amssymb}
\RequirePackage[authoryear]{natbib}
\usepackage[utf8]{inputenx}
\usepackage{indentfirst}
\usepackage{microtype}
\usepackage{comment}
\usepackage{epigraph}
\usepackage{url}
\usepackage{tikz}
\usepackage{makecell}
\usepackage{hyperref}
\usepackage{listings}
\usepackage{lstbayes}

\usepackage{tabularx}
\usepackage{booktabs}
\usepackage{multicol}
\usepackage{multirow}

\usepackage{caption,subcaption}
\usepackage{float}
\usepackage{graphicx}
\usepackage{rotating}
\usepackage{color, colortbl}

\usepackage{mathtools}
\usepackage{mathbbol}
\usepackage{amsbsy}

\usepackage{amsfonts}
\usepackage{soul}
\usepackage{bbm}
\usepackage{bm}

\newcommand{\bolds}[1]{\boldsymbol{#1}}

\newcommand{\calS}{{\cal S}}
\newcommand{\calT}{{\cal T}}

\newcommand{\bD}{\bolds{D}}

\newcommand{\bbE}{\mathbb{E}}

\newcommand{\bI}{\bolds{I}}

\newcommand{\bQ}{\bolds{Q}}

\newcommand{\bv}{\bolds{v}}
\newcommand{\bV}{\bolds{V}}

\newcommand{\bw}{\bolds{w}}
\newcommand{\bW}{\bolds{W}}
\newcommand{\bx}{\bolds{x}}
\newcommand{\bX}{\bolds{X}}
\newcommand{\by}{\bolds{y}}
\newcommand{\bY}{\bolds{Y}}

\newcommand{\IdentityMat}{\bI}
\newcommand{\Norm}{\mathcal{N}}
\newcommand{\bzero}{\mathbf{0}}

\newcommand{\bbeta}{\bolds{\beta}}

\newcommand{\btheta}{\bolds{\theta}}

\newcommand{\bpsi}{\bolds{\psi}}
\newcommand{\bPsi}{\bolds{\Psi}}
\newcommand{\bphi}{\bolds{\phi}}

\newcommand{\given}{\,|\,}
\newcommand{\lrnd}{\left(}
\newcommand{\rrnd}{\right)}
\newcommand{\lsq}{\left[}
\newcommand{\rsq}{\right]}
\newcommand{\lcur}{\left\lbrace}
\newcommand{\rcur}{\right\rbrace}
\renewcommand{\tilde}{\widetilde}

\title{A Bayesian spatio-temporal extension to Poisson Auto-Regression: modeling the disease infection rate of Covid-19 in England
}

\author{
  Pierfrancesco Alaimo Di Loro\\
  Department GEPLI \\
  LUMSA University \\
  \texttt{p.alaimodiloro@lumsa.it} \\
   \And
  Dankmar B\"ohning, Sujit Sahu \\
  University of Southampton \\
  Mathematical Sciences and Social Statistics Research Institute \\
  \texttt{xq1u19@southamptonalumni.ac.uk, d.a.Bohning@soton.ac.uk, s.k.sahu@soton.ac.uk} \\
}

\begin{document}
\maketitle

\begin{abstract}
The COVID-19 pandemic provided many modeling challenges to investigate the evolution of an epidemic process over areal units. A suitable encompassing model must describe the spatio-temporal variations of the disease infection rate of multiple areal processes while adjusting for local and global inputs. We develop an extension to Poisson Auto-Regression that incorporates spatio-temporal dependence to characterize the local dynamics while borrowing information among adjacent areas. The specification includes up to two sets of space-time random effects to capture the spatio-temporal dependence and a linear predictor depending on an arbitrary set of covariates.
The proposed model, adopted in a fully Bayesian framework and implemented through a novel sparse-matrix representation in Stan, provides a framework for evaluating local policy changes over the whole spatial and temporal domain of the study. It has been validated through a substantial simulation study and applied to the weekly COVID-19 cases observed in the English local authority districts between May 2020 and March 2021. The model detects substantial spatial and temporal heterogeneity and allows a full evaluation of the impact of two alternative sets of covariates: the level of local restrictions in place and the value of the Google Mobility Indices. The paper also formalizes various novel model-based investigation methods for assessing additional aspects of disease epidemiology. 
\end{abstract}


\section{Introduction}
\label{sec:intro}

COVID-19 is the infectious disease caused by the virus SARS-COV-2, first identified in Wuhan (China) in December 2019. Attempts to contain the initial Chinese outbreak failed and the virus was able to spread worldwide in record time. The \textit{World Health Organization} (WHO) listed the COVID epidemic as a public health emergency of international concern on the 30th of January 2020 and it was upgraded to a global pandemic a couple of months later.
The advent of the COVID-19 pandemic caught many governments unprepared and disrupted the lives of millions of people around the world. The need to safeguard public health and National health systems called for urgent \textit{``Non-Pharmaceutical Interventions''} (NPIs) by all National authorities. 

In England, the first COVID-19 case was detected on February 28, 2020. The government had to take immediate action and initially advocated for mild containment measures to favour \textit{herd immunity}. However, it decided to change strategy as soon as the first modelling results on the spread of the disease were available. These initial modelling attempts, whilst extremely rough, highlighted how tougher restrictions were needed to avoid a high death toll and to protect the sustainability of the public National Health System (NHS).
At the start of the pandemic, very little was known about the various characteristics of the virus and all governments operated in a context of great uncertainty. Data collection systems for monitoring the spread were immediately set up but took some time to function fully. These delays slowed down the epidemiological research on the virus and hampered the explanation and forecasting of its infection dynamics. These aspects are key to supporting the government's decision-making process on any public health interventions, most especially in a delicate situation such as the COVID-19 pandemic.

The most natural solution to model epidemiological data is represented by compartmental models, see e.g. \cite{diak:13, ferguson2020imperial, danon2021spatial}. These models provide a mechanistic description of the contagion dynamic of the disease through differential equations. In principle, they allow us to determine the current reproduction rate of the virus $R_t$, i.e. the average number of people infected by a single infected individual during a given time window ($t$). Estimation of $R_t$ relies on the knowledge of the system's initial conditions (e.g. current fraction of infected and susceptible individuals, etc.) and accurate preliminary estimates of several virus characteristics (e.g. the basic reproduction rate, the incubation period, the serial interval, etc.).
While such quantities may be well-known for widely researched viruses and diseases, their reliable estimation for a new one requires time and statistically valid data collected through well-designed epidemiological studies. This has proved to be particularly challenging in the context of the COVID-19 pandemic.
First of all, when an epidemiological emergency is in place the surveillance conditions change as the epidemiological process evolves. The data cannot be gathered under a stable collection scheme because of the continuous public health authorities' interventions and population behavioural adjustments.
Second, like other Coronaviruses, SARS-Cov-2 is extremely prone to mutations that can significantly alter its defining characteristics. Such properties of the virus introduce a large amount of random and non-random variation in the collected data, which results in great uncertainty of the outcome and of all the derived quantities \citep{chowell2017fitting, baek2021limits, zhao2021}. In particular, \cite{Ioannidis2020} shows how poor data input on the above-mentioned key features can heavily bias the estimates of compartmental models, jeopardizing the reliability of any theory-based forecasting effort.

To investigate alternatives to compartmental models, much research has been devoted to the development of empirical models that sacrifice pertinence in favour of more robust inferences \citep{leah2018, salje2020estimating, farcomeni2020ensemble}. This approach deals with the data flow at the macroscopic level and adopts statistical modelling to describe multiple aspects of the epidemic from a global perspective \citep{chowell2016using, harvey2020time, alaimo2021nowcasting}. All inferences are drawn from the statistical properties of the observed data: long and short-term trends, dependence patterns, and observed and unobserved heterogeneity. This allows full quantification of the uncertainty without imposing any strong assumption or restriction on the process behaviour. 

The main aim of this paper is to develop a novel Bayesian spatio-temporal model for the disease infection rate as an extension to the general Poisson auto-regression framework \citep{kedem2005regression, fokianos2009poisson, xu2020, Giudici2022}.
At the top level, the model assumes that the weekly cases are the realizations of Poisson random variables where the infection rate is decomposed into the sum of an auto-regressive term and a baseline. The first is named the \textit{epidemic part}, whilst the second is the \textit{endemic part} (endemic-epidemic model, \cite{held2005statistical, HeldBio2006}).
The auto-regressive coefficient of the epidemic component captures the inherent dependence of each count on the previous ones. It determines the \textit{epidemic growth rate of detected cases} $\tilde{r}$, which is just a portion of the \textit{global epidemic growth rate} $r$ \citep{jewell2021use, parag2022}. 
This distinction induces a major source of heterogeneity in the data, as the cases detected at each time cannot be explained only through the reproduction of the cases detected at previous times. The residual component is the result of a hidden process brought on by the mass of undetected cases, which is captured by the baseline.

This mechanism has been adopted by many authors \citep{agosto2020poisson, agosto2021monitoring, Giudici2022} to model the COVID-19 epidemic at the aggregate national level and none of them considers the effect of external covariates nor induces any sort of dependence across units. 
Other authors induce direct dependence of infected counts across nearby areas by adopting a multivariate version of the endemic-epidemic model \citep{knorr2003hierarchical, held2012modeling, celani2022endemic} but do not account for any sort of spatio-temporal variability in the parameters.
However, the (relative) magnitudes of the epidemic and endemic components depend on multiple observed and unobserved factors that vary over space and time in a possibly smooth way.

To fill a gap in the literature,  we propose a unified spatio-temporal model for the infection rates of multiple areas that extends the endemic-epidemic model with static and/or non-spatially varying growth rates whilst accounting for cross-dependence. We are not aware of a spatio-temporal model for the disease infection rate, or the popularly known disease reproduction number $r$. In effect, in the Poisson-autoregression model,  we propose to use $r(\bolds, t)$ where $\bolds$ indexes space and $t$ indexes time instead of a single number $r$ for a spatial or temporal or spatio-temporal domain.   
We assume a spatially dis-aggregated model for area-wise disease growth, but at the same time borrow information in space and time by adopting methods from the domain of spatial \cite{gelfand2010handbook} and spatio-temporal disease mapping \citep{knorr2000bayesian, bohning2000space}. The latter has seen prime attention in recently published computer software packages \citep{sahu2022bayesian} and it is the basis of many articles on COVID-19 cases \citep{jalilian2021hierarchical, bartolucci2021spatio, mingione2022spatio}, where specific applications to the UK regions include \cite{mishra2020covid, konstantinoudis2021long, sahu2022bayesianPaper}. 
We inject space-time variation into the two main components defining this endemic-epidemic model: the auto-regressive coefficient $\tilde{r}\rightarrow \tilde{r}_{\ell t}$ and the baseline $b\rightarrow b_{\ell t}$, with $(\ell, t)\in \calS \times \calT$ being the whole spatio-temporal domain.
We achieve that by introducing two sets of random effects with non-separable space-time dependence as induced by the space-time Leroux model \citep{rushworth2014spatio}.
Our specification implies smooth variations of the coefficients in space and time and allow the auto-regressive coefficient to describe both the expanding ($\tilde{r}_{\ell t}>1$) and the contracting ($\tilde{r}_{\ell t}<1$) phases of the epidemic.
Furthermore, the pooling of all areas in a joint model yields more robust inferences on the components governing the single epidemic processes, which is an extremely relevant aspect when dealing with areas of modest size that share common characteristics.

We apply our model to the weekly series of new positive cases detected in the English \textit{''Local Authorities, Districts, Counties, and Unitary Authorities``} (LADCUA) during the first year of the pandemic from April 2020 to March 2021. Each LADCUA has partial health autonomy from the central government and England has been one of the first nations to implement local policies to contain spatially delimited outbreaks.
The segmented policy decisions implemented through the tiering system make it clear that the epidemiological process cannot be homogeneous across all of England over the entire time window. It is also well-established how the aggregated rate, and hence the National growth rate, is not representative of any of the rates affecting the single areas \citep{burghardt2022unequal}. It is then crucial to keep that into account by embedding in the model the topological feature of the districts, i.e., their mutual arrangement and connections, their spatiality.
Thus, we consider alternative specifications of the proposed model (from the least to the most complex) to scrutinize for the presence of possible exogenous effects of several spatial and time-varying covariates. In particular, we focus on indicator variables denoting whether a LADCUA was placed in tier I, II, III or IV during the previous week. Alternatively, we consider variables taken from the \textit{Google Community Mobility Reports} \citep{googlecr}, that indicate the extent of people aggregation in different kinds of areas (e.g. residential, retail and recreation, transit stations, etc.) in each LADCUA.

The proposed model enables the investigation of several aspects of the disease dynamic, both from a global and a local perspective. First, we are able to estimate the indirect effects of tiering-based restrictions on the spread of the epidemic in a coherent, dynamic, and joint framework. Several other authors embarked on the same mission \citep{flaxman2020estimating, davies2021association, laydon2021modelling, pelagatti2021assessing, zhang2022evaluating}, but none of these developed a comprehensive framework that describes the underlying epidemiological process and properly propagates the uncertainty at all levels. Secondly, we are able to investigate whether the dynamic is better described by the \textit{Google Mobility Indices} (GMI) than by the tiering system. In particular, we are able to identify areas that are estimated to be more vulnerable to the spreading of cases.
Finally, another interesting novel by-product of our model is the possibility of quantifying the weekly number of cases that can be explained through the contagions of the previous weeks and those numbers that are instead caused by undetected infections.  


The remainder of the paper is organized as follows.
Section~\ref{sec:data} introduces the data of interest explaining the adopted data collection and the cleaning process.
This section also presents interesting summary statistics and exploratory graphical displays to better understand various features of the compiled data set. Section~\ref{sec:model} develops the proposed model and within it Section~\ref{subsec:STAN} provides details of model implementation in STAN \citep{carpenter2017stan, Stan, Rstan}. 
The application is contained in Section \ref{sec:app}, which includes a simulation study and the final analysis results in Sections \ref{subsec:sim} and \ref{subsec:res}, respectively.
Section \ref{sec:conc} contains some concluding remarks and considerations for further developments.

\section{The data}
\label{sec:data}

The spatial domain of our analysis includes the nine regions in England\footnote{Wales, Scotland and Northern Ireland have been excluded as they have developed different health policies and interventions.}: London, North-East, North-West, East Midlands, West Midlands, Yorkshire, South-East, East of England, and South-West. The basic areal spatial units are the 313 LADCUAs as they were designed by area names, codes, and boundary maps  provided  the UK Office of National Statistics (ONS) in May 2020. These LADCUAs include the 32 boroughs in Greater London, but excludes the one for Isle of Scilly in the Atlantic ocean off the coast of Cornwall. Details about the data collection process are provided in the Supplementary Material.

Our data set comprises of the weekly number of positive cases in each of the 313 English LADCUAs between April 4, 2020 and March 8, 2021. This time window covers the second and third waves of the COVID-19 pandemic in England. The starting date (April 4, 2020) is the first date for which the figures related to all LADCUAs are available, while the 8th of March 2021 signs the end of the third (and last) National lockdown. Data after 8th of March 2021  have not been included in the analysis because of  many sudden changes which impacted the epidemic process in very short time-windows. For example,  the vaccination campaign was operating fully and started the administration of both the required first and second doses; the use of \textit{Lateral-Flow-Device} (LFD) testing was becoming more common place than \textit{Polymerase Chain Reaction} (molecular, PCR) testing; the Delta variant would shortly hit the UK and boost the transmission and detection process of the virus. Henceforth we limit our attention to analyzing  the data in the stated time window  to evaluate the impact of National and Local Policies, i.e. the tiering system. Such policies have been in action from July 2020 (the first Leicester local lockdown) to January 2021 (the start of the last National lockdown).

\begin{figure}[tbp]
    \centering
    \includegraphics[width=0.8\textwidth]{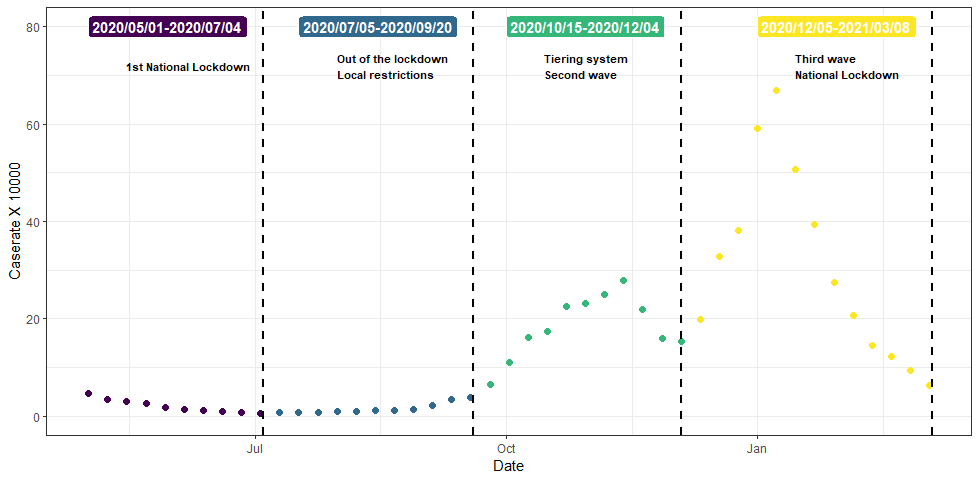}
    \caption{Weekly case rate per 1000 inhabitants in England between May 2020 and March 2021}
    \label{fig:timeline}
\end{figure}

Let the weekly case-rate of be the total number of new cases detected in an area during a specific week divided by the corresponding population size. Figure \ref{fig:timeline} shows a timeline of the overall English weekly case-rate between May 2020 and March 2021. The figure highlights some important passages of the analyzed time-period, identifying four main windows. In order: the final tail of the first National Lockdown is window 1; exit from the first National lockdown and the implementation of local targeted restriction policies is window 2; the beginning of the second wave, the implementation of the first tiering system, and the second National lockdown fall in window 3; the beginning of the third wave, the introduction of the new tiering system, and the third National Lockdown are in window 4.

\begin{figure}
    \centering
    \includegraphics[trim={3cm 0 3cm 0}, clip, width=0.9\linewidth]{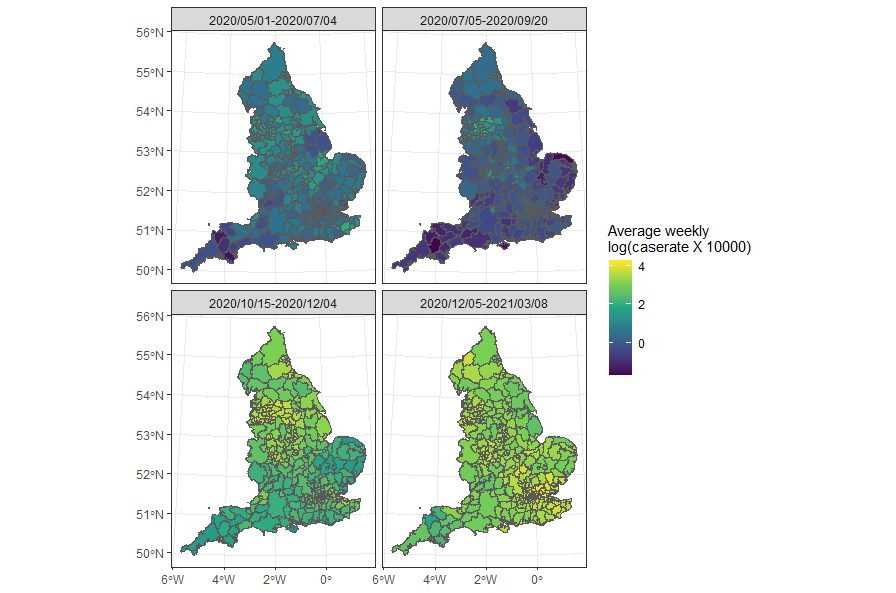}
    \caption{Average weekly case rate per 1000 inhabitants in the 313 English LADCUAS withing the four relevant windows between May 2020 and March 2021}
    \label{fig:MapsCases}
    
\end{figure}

Figure \ref{fig:MapsCases} shows the average weekly case-rate observed in each district throughout the four above-mentioned windows. It is represented on the log-scale to illuminate the spatial differences within each of the four windows. 
We can see a clear temporal gradient, already noticeable from the timeline in Figure \ref{fig:timeline}, with the first two windows having way lower case-rates than the last two. 
We notice how the pattern is pretty uniform all over England during the first window. A cluster of low-rates districts in the south-west is an exception, and it will preserve this characteristics in all windows.
Whilst still low, we can notice an increase in the case-rates of the northern regions during the second window (top-right). The Liverpool and Manchester areas in particular. 
The second wave (third window, bottom-left) hit the north, with the north-west especially affected. The southern regions are instead less interested by this wave. 
Finally, the last window (i.e. the third wave) hit all over England. The London region seems to have been particularly affected by this last wave.

This preliminary visual inspection brings out the substantial spatial clustering in the data, stressing the need to deal with the spatially structured heterogeneity in the data.

\subsection{Socio-economic variables}
It is reasonable to think that difference in socio-economic factors among areas can  impact on the spread potential of the virus. Whilst there is no final proof yet, there is an ongoing debate on how wealth differences can impact the risk of contracting the virus \citep{gangemi2020rich, gong2022wealth, Padellini2022}. 
Following \cite{sahu2022bayesian}, we consider the following small but relevant set of space-varying (but time-constant) socio-economic variables.

\begin{itemize}
    \item The \textit{Population size}, which is key to offset the different demographic dimensions of the various LADCUAs.
    \item The \textit{Population density}, as a larger value of it could imply a greater tendency to aggregate and therefore favor the spread of the virus.
    \item The percentage of the working age population receiving job-seekers allowance during January 2020 (denoted \textit{jsa}).
    \item The median house price in the district in March 2020 (denoted \textit{hprice}). 
\end{itemize}

We are mostly interested in understanding how these variables affect the growth rate of the epidemic in the various LADCUAs.

\subsection{Local policies}
England has been one of the first countries to implement \textit{Non-Pharmaceutical Interventions} (NPIs) at the local level to contain localized outbreaks. The first example dates back to July 2020, when only the Leicester area was put under a hard lockdown to contain an uncontrolled outbreak of cases. Many other areas have followed these lockdowns subsequently. From October 2020 to January 2021 the local policies were \textit{institutionalized} through the adoption of a tiering system based on $3$ different degrees of alert levels (Tiers I,  II, and III). An additional Tier IV, akin to the total lockdown, has been introduced only to face the third wave during Christmas time in 2020.
All the implemented local policies, area by area, are  available from the government web pages\footnote{\url{https://www.gov.uk/government/collections}; \url{https://www.gov.uk/guidance}}. We reconstructed the timeline of local restrictions through different sources. Details about their characterization and backdating to times when a more destructured system of local policies was in place are included in the Supplementary Material.

\begin{figure}[tbp]
    \centering
    \includegraphics[width=0.9\textwidth]{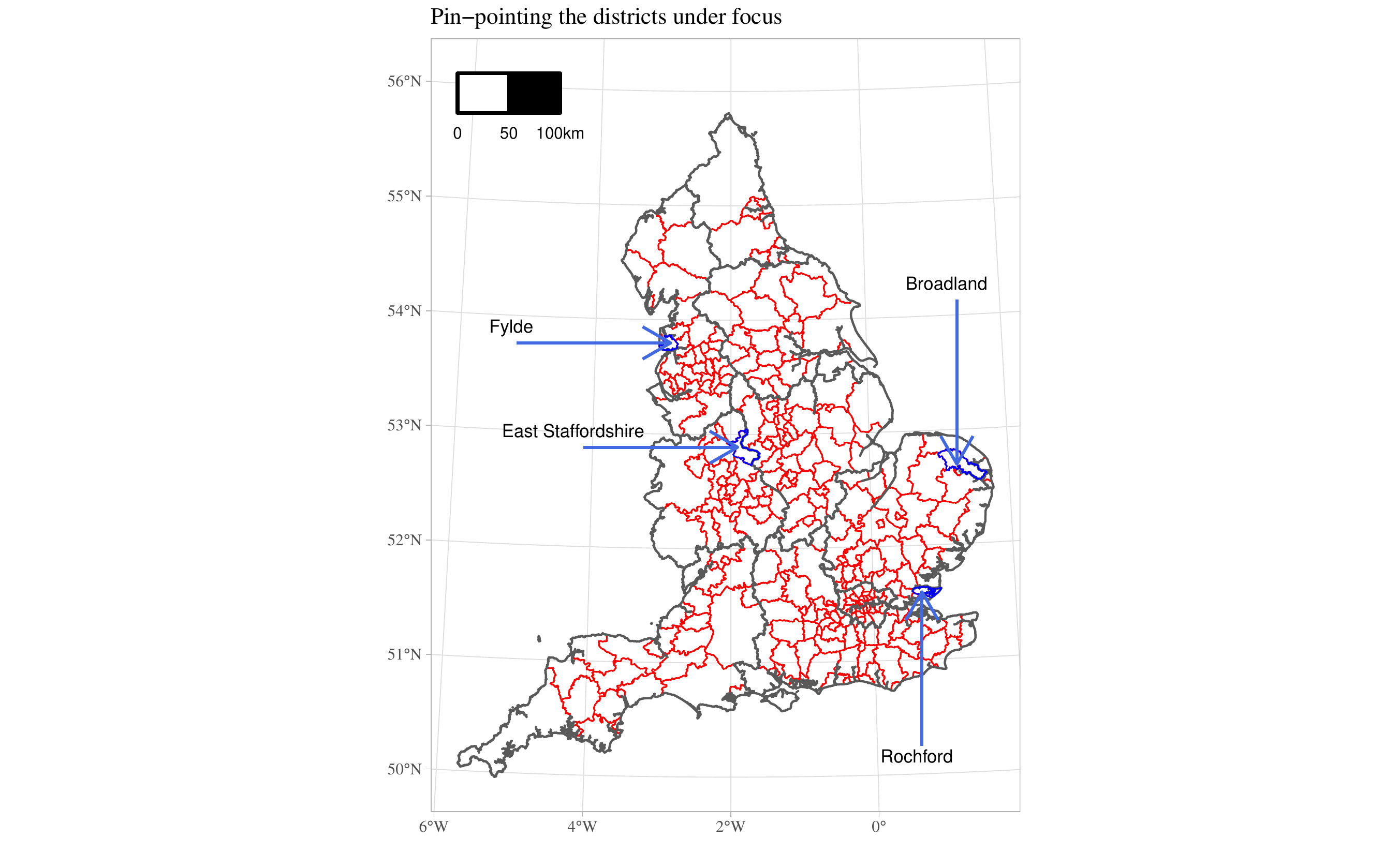}
    \caption{Location of the randomly selected authorities of Broadland, East Staffordshire, Fylde, and Rochford.}
    \label{fig:pinpoint}
\end{figure}

Figure \ref{fig:caseratel} shows the time-series of the weekly case-rate in the four randomly selected authorities of Broadland, East Staffordshire, Fylde, and Rochford, color-coded by the corresponding tier restriction. Their location is pin-pointed in Figure \ref{fig:pinpoint}. The figure clearly shows how higher level restrictions usually correspond to contracting phases of the epidemic, and vice versa. Nevertheless, this is not an exact rule as there are many other unobserved factors that govern the epidemic process. For instance, it is apparent how Tier III local restrictions were more effective in some cases and less effective in others. 

\begin{figure}[tbp]
    \centering
    \includegraphics[width=0.9\textwidth]{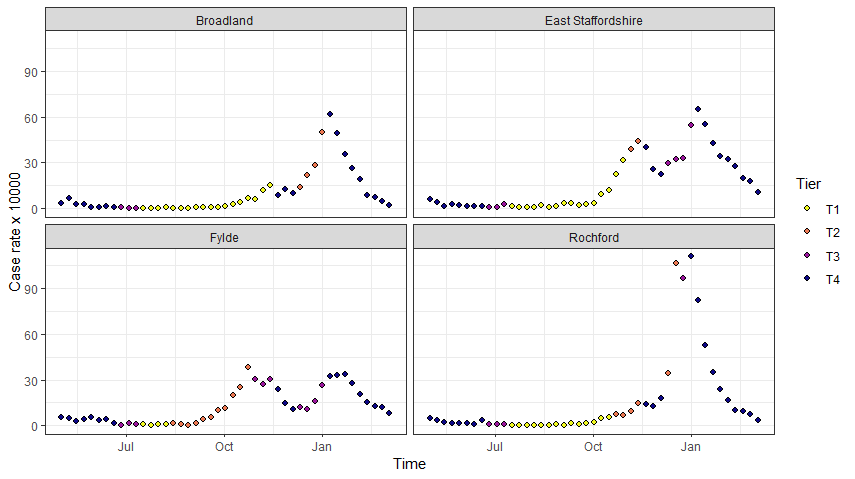}
    \caption{Time-series of the weekly case-rate X 10000 inhabitants in the randomly selected authorities of Broadland, Colchester, Darlington, East Staffordshire, Fylde, Rochford.}
    \label{fig:caseratel}
\end{figure}

We also  observe how the epidemic did not grow during the first part of Tier I restrictions, that was in place  from July to September 2020. This behavior can be explained through a multitude of factors impacting the epidemic process during the summer time: people spend longer time outdoors because of the more favorable weather; higher temperatures seem to have a negative effect on the virus spreading potential, see e.g. \cite{ganslmeier2021impact}; schools close for the Summer vacations, eliminating one of the vectors through which respiratory diseases spreads among different households. 
At the same time, we  notice a sudden acceleration of the epidemic process in December 2020, which may by explained through the social gathering that normally characterize the Christmas time.
We may introduce two additional indicator variables to account for such regime change, but defining the exact application window would be cumbersome. 

It is quite complex to define and delimit these two effects, as they also mix with other external factors (e.g. scarcity of tests in mid-September 2020, increase in testing requests in early December 2020), and it is a more prudent choice to let this unspecified and see if the random affects are able to detect this pattern unassisted


\subsection{Google community mobility report - GMI}
The adoption of non-pharmaceutical Interventions (NPIs, e.g. the tiering system) has an indirect effect on the spreading of the virus. It regulates people's  behavior and habits, which in turn  determines the contagion and the epidemic growth rate.
The naive tiers level are only a coarse approximation to the actual changes in the behavior of the population when responding to such restrictions. This is due to multiple reasons: delays in the effects of the implemented policies, people willingness to respect the rules (which is variable in space and time), the degree of enforcement implemented by the local authority, etc.
The \textit{"Google community mobility report"} provides an easy and direct way to evaluate people movement and aggregation trends across different categories of places.
Places are organized in six different categories: \textit{Grocery and Pharmacy}, \textit{Parks}, \textit{Transit stations}, \textit{Retail and recreation}, \textit{Workplaces}, and \textit{Residential}.
These data show how visits and length of stay at different places change (in percentage) compared to a baseline. The baseline is the median value for the corresponding day of the week in the same area during the 5-weeks period that goes from January 3 and February 6, 2020.

A more detailed description of the collection process of these indices and their behavior with respect to the tiering system are included in the Supplementary Material

\section{The model: formalization and implementation}
\label{sec:model}

We observe the epidemic process over $t=1,\dots, T$ times and across the $L$ regions in $\calS=\lcur \boldsymbol{\ell}_1,\dots,\boldsymbol{\ell}_L\rcur$.
Let $\bY=\lsq\by_1,\dots,\by_T\rsq$ be the $L\times T$ matrix of the $n=L\cdot T$ observed cases, with $\by_t=\lsq y_{1 t},\dots,y_{L t}\rsq$ the spatial vector at time $t$.
We consider the Poisson auto-regression to account for the count and epidemiological nature of the data. As highlighted in the body of Section \ref{sec:intro}, it allows to disentangle the process into two components: an autoregression directly depending on the past counts and a baseline independent from past counts.
Counts at each $\ell\in\calS$ and time $t$ depend directly on the counts of the previous time as:
    \begin{equation}
    \label{eq:poiAR1}
    \begin{aligned}
    &Y_{\ell t}\,|\,\by_{\ell (1:t-1)}\sim Poi(\lambda_{\ell t}),\\
    &\lambda_{\ell t} =  \tilde{r}_{\ell}\cdot y_{\ell (t-1)}+ b_{\ell},
    \end{aligned}
\end{equation}
where $\tilde{r}_{\ell}$ and $b_{\ell}$ are the location-specific auto-regressive coefficients and a baseline intercept, respectively.
The former determines the memory of the process, regulating the impact of the previous count on the current one. From an epidemiological perspective, $\tilde{r}_{\ell}$ is the \textit{epidemic growth rate of the detected cases} and it is the main component driving if and how fast the epidemic will grow or decline. Notice that the process is stationary only for $\tilde{r}_{\ell}<1$.  
The latter is a baseline independent from the previous count. This term avoids the process to collapse to $0$ if, at any time, a $0$ count ($y_{\ell (t-1)}=0$) is observed. It is necessary to account for the effect of undetected, imported cases, or spontaneous cases that will keep the virus circulation alive even if no cases are detected at previous times. It ensures that the process will not die out with probability $1$ \citep{held2005statistical}.


This full specification empowers this model to flexibly describe any type of real-life epidemic process where
the spread of the disease displays a mixture  of both endemic and epidemic behaviour \citep{hautsch2011econometrics, matteson2011forecasting, gross2013predicting, zhu2016local, agosto2016modeling, xu2020}, and have been recently used to model the epidemic processes of COVID-19 and other diseases \citep{struchiner2015increasing, agosto2020poisson, agosto2021monitoring, chan2021count, celani2022endemic}.
A major limitation of these methods is that they usually assume the time-series as stationary and with time-constant coefficients. However, the epidemiological process of the COVID-19 epidemic is not stationary. Since its explosion, it alternated between expanding ($r_{\ell}>1$) and contracting ($r_{\ell}<1$) phases as a consequence of NPIs, testing policy changes, behavioural adjustments in the population, and so on. This calls for extending the model to have time-varying parameters $\tilde{r}_{\ell t}$ and $b_{\ell t}$, with the growth rate of detected cases $\tilde{r}_{\ell t}$ that can eventually be larger than $1$ and imply non-stationary periods in the process.
Furthermore, most modelling efforts of this type have been focused on modelling nationally aggregated counts \citep{struchiner2015increasing, agosto2020poisson, Giudici2022}. In such cases, the parameters relating to the epidemics in different countries are estimated independently. Here, we focus on small areas that constitute a single Nation such as England. Therefore, it is reasonable to assume that there is substantial spatial dependence between the local processes, especially among neighbouring districts. 
This dependence can be attained by introducing a set of random effects that borrow information through space and time but keep the necessary flexibility to capture any time and district-specific variation.


Furthermore, the specification in Equation \eqref{eq:poiAR1} can be enriched by envisioning the possibility that a certain part of the observed count at time $t$ can be explained through counts at lags larger than one. We do so by applying the auto-regressive term on a weighted average of the previous counts up to a pre-specified lag $\tau$ so that the cases of each week deplete their full effect in the following $\tau$ weeks.
Finally, we also account for the fact that an epidemic cannot grow indefinitely. Its growth is limited in size by the population it is spreading within, as already infected people cannot be reinfected within a certain time window. There is an ongoing discussion on how long such immunity shall last for the SARS-Cov-2. Substantial evidence shows that during the first year of the pandemic before the Delta and Omicron variant hit, the probability of re-infection has been practically null \citep{ren2022reinfection, o2022quantifying}.
Settling for such precautionary value, we can account for the reduced number of individuals exposed to infection in each district at each time by re-scaling the corresponding rate by:
\begin{equation*}
    d_{\ell t} = \frac{\sum_{j=1}^{\tilde{t}} y_{\ell (t-j)}}{pop_{\ell}},\qquad \forall\, \ell\in\calS,\; t=1,\dots, T,
\end{equation*}
where $pop_{\ell}$ is the population size of district $\ell$.

Thus, the rate at location $\ell\in\calS$ and time $t=1,\dots,T$ can be expressed as:
\begin{equation}
\label{eq:poiAR}
        \lambda_{\ell t} = \lrnd\sum_{i=1}^{\tau} \lrnd w_i\cdot y_{\ell (t-i)}\rrnd \cdot \tilde{r}_{\ell t} + b_{\ell t}\rrnd\cdot d_{\ell t}
        ,
\end{equation}
with $\sum_{i=1}^\tau w_i=1$, $\tilde{r}_{\ell t}>0$ and $b_{\ell t}>0$ for all $\ell\in\calS$ and $t\in\lcur 1,\dots,T\rcur$.
Our main interest lies in modelling the auto-regressive coefficient, which could be expressed as a function of covariates and random effects through the \textit{log-link} function:
\begin{equation}
\label{eq:epGR}
        \log\lrnd\tilde{r}_{\ell t}\rrnd= \bx_{\ell t}^\top\cdot\bbeta + \phi_{\ell t},
\end{equation}
where $\bx_{\ell t}$ is a $(k+1)\times 1$ vector of (potentially space-time varying) covariates and intercept, $\bbeta=\lsq\beta_0,\dots,\beta_k\rsq$ is a vector of $k+1$ coefficients, and $\lcur\phi_{\ell t}\rcur_{\ell t}^{LT}$ is a set of space-time correlated random effects. The intercept $\beta_0$ determines the average (over space and time) growth rate of the epidemic, which would be $r_0=e^{\beta_0}$.
Very similarly, we can model the baseline on the log-scale as well. To make it scale-invariant it shall depend on a location-specific offset $\text{off}_\ell, \ell\in\calS$:
 \begin{equation}
 \label{eq:bGR}
\log\lrnd b_{\ell t}\rrnd=\log\lrnd\text{off}_{\ell }\rrnd+\bv_{\ell t}^\top\cdot\bolds{\eta}+\psi_{\ell t},
\end{equation}
where $\bv_{\ell t}$ is a $(\nu+1)\times 1$ vector of (potentially space-time varying) covariates, $\bolds{\eta}=\lsq\eta_0,\dots,\eta_\nu\rsq$ is the corresponding vector of coefficients, and $\psi_{\ell t}$ is a second set of space-time correlated random effects  affecting the baseline. The latter allows explaining any extra-variability in the process and accounts for over-dispersion with respect to the Poisson assumption.

The specification of the auto-regressive coefficient in Equation \eqref{eq:epGR} seem to lack a scaling factor that accounts for the size of the population the virus is spreading in. When dealing with heterogeneous spatial units this can be a source of substantial bias, as smaller regions have less growth potential than larger ones.
Introducing the population size as an offset is not entirely satisfactory as the number of personal contacts (hence the potential reproduction of detected cases) does not necessarily depend on the population size.
On the other hand, the global scaling factor $d_{\ell t}$ indirectly controls for the population size by tempering the growth in smaller regions. Indeed, given the same number of past observed cases in two different regions, the one with the smaller population will have a smaller $d_{\ell t}$ than the other.

\subsection{A CAR-AR prior for the space-time random effects}
There is a rich literature on \textit{Gaussian Markov Random Fields} and \textit{Conditional Auto Regressive} (CAR) priors \citep{besag1974spatial, rue2005gaussian} that suit the modelling of dependent random effects over a network. They have seen wide applicability in the disease mapping context \citep{lawson2018bayesian}, and much interest has been recently devoted to their extension in the space-time setting \citep{lee2020estimation, sahu2021bmstdr}.
We here consider the space-time extension of the Leroux model, see \cite{leroux2000estimation}, originally proposed by \cite{rushworth2014spatio}, that we denote as \textit{CAR-AR Leroux}.

Let $\phi_{\ell t}, \;\ell\in\calS,\, t=1,\dots,T$ be the set of \textit{Gaussian} space-time random effects over the discrete domain $\calS=\lcur \boldsymbol{\ell}_1, \dots, \boldsymbol{\ell}_L\rcur$ introduced in Equation \ref{eq:epGR}. Here, we will specify everything in terms of this set of random effects, but the same reasoning holds for $\bPsi=\lsq\bpsi_1,\dots,\bpsi_T\rsq$ as is. 

Let $\bW$ be a $L\times L$ adjacency matrix for the locations in $\calS$, whose elements $w_{ij}$ are equal to $1$ if and only if $i\neq j$ and $i\sim j$ (location $\boldsymbol{\ell}_i$ is a neighbor of location $\boldsymbol{\ell}_j$), and $0$ otherwise. Notice that each location is not a neighbor to itself. The number of neighbors of each location $\boldsymbol{\ell}_i$ is $N_i=\sum_{j=1}^lw_{ij}$.

Slicing through time, the original \textit{Intrinsic Auto-Regressive} (ICAR) specification of \cite{besag1974spatial} expresses the prior \textit{conditional} mean of each element $\phi_{it}$ as the simple average of its neighbors $\bbE\lsq \phi_{it}|\bphi_{t}\rsq=\frac{1}{N_i}\sum_{j\sim i}\phi_{jt}$. This corresponds to the following \textit{improper} joint specification:
\begin{equation*}
    f(\bphi_t)\propto \exp\lcur -\frac{1}{2}\sum_{i\sim j}(\phi_{it}-\phi_{jt})^2\rcur= \exp\lcur -\frac{1}{2}\cdot\bphi_t^{\top}\lrnd\bD-\bW\rrnd\bphi_t\rcur,
\end{equation*}
where $\bD$ is the diagonal matrix with entry $d_{ii}$ equal to the number of neighbors of unit $i$ and $\bphi_t=\lsq\phi_{1t},\dots,\phi_{Lt}\rsq$

\cite{leroux2000estimation}, among others, extends the model to include the \textit{spatial smoothing} coefficient $\alpha\in (0, 1)$. It regulates the extent of the spatial dependence and makes the joint prior proper:
\begin{equation}
\label{eq:Leroux}
    \bphi_t\sim \Norm_L\lrnd \bzero,\;\sigma^2\cdot\bQ(\alpha,\,\bW)\rrnd,
\end{equation}
where $\bQ(\alpha,\,\bW)=\lrnd \alpha\cdot(\bD-\bW) + (1-\alpha)\IdentityMat_L\rrnd$ and $\sigma^2>0$.
This yields the conditional mean $\bbE\lsq \phi_{it}|\bphi_{t}\rsq=\frac{\alpha}{N_i+1-\alpha}\sum_{j\sim i}\phi_{jt}$ and recovers the complete independence and ICARs settings for $\alpha=0$ and $\alpha=1$, respectively.

The space-time extension by \cite{rushworth2014spatio, rushworth2017adaptive} connects the $T$ time slices through a first-order auto-regressive structure:
\begin{equation}
\label{eq:LerouxAR}
    \begin{aligned}
        &\bphi_1\sim \Norm_L\lrnd \bzero,\, \sigma^2\cdot\bQ\lrnd\alpha, \bW\rrnd^{-1}\rrnd\\
        &\bphi_t\given\bphi_{t-1},\dots,\bphi_1\sim \Norm_L\lrnd \rho\cdot\bphi_{t-1},\, \sigma^2\cdot\bQ\lrnd\alpha, \bW\rrnd^{-1}\rrnd,\; t=2,\dots,T,      
    \end{aligned}
\end{equation}
where $0<\rho<1$ is the temporal auto-regressive coefficient and regulates the amount of temporal dependence. We denote with $\btheta=(\alpha, \rho, \sigma)$ the vector of coefficients on which the CAR-AR specification depends on.

\subsection{Sparse implementation in STAN}
\label{subsec:STAN}
The model has been implemented in \texttt{R} via STAN \citep{Stan, Rstan} that uses the Hamiltonian Monte Carlo routine known as No-U-Turn sampler \citep{hoffman2014no}.
It is a very general-purpose software for MCMC estimation, that provides many diagnostic tools that can detect identifiability or miss-specification issues in the model. 

The implementation of the Leroux model for large networks ($L>100$), in the space-time setting especially, is computationally expensive. Indeed, the evaluation of the density in \ref{eq:Leroux} encompasses the computation of a quadratic form on the $L\times L$ matrix $Q(\alpha,\bW)$ and its determinant.
This bottleneck is typical of many other CAR specifications, among which the \textit{proper CAR} \citep{cressie2015statistics}, and it requires the adoption of ad-hoc strategies to estimate the model in a reasonable amount of time.
First, we can notice that only a few entries of $\bW$, and hence of $\bQ(\alpha, \bW)$, are non-zero. This can significantly speed up the evaluation of the quadratic form if we exploit a sparse-representation of $\bW$ in the corresponding algebraic operations.

Second, we draw from \cite{jin2005generalized} who suggested an efficient computational strategy to evaluate the determinant in the \textit{proper CAR} setting.
We here, for the first time, extend that same strategy onto the Leroux model by rewriting the precision matrix to have a comparable structure. Indeed:
\begin{equation}
    \label{eq:effPrec}
        Q(\alpha,\bW)\, = \, \lrnd\IdentityMat_l-\alpha\lrnd\bW+\IdentityMat_l-\bD\rrnd\rrnd .
\end{equation}
Now, let $\lambda_i=1,\dots, l$ be the eigenvalues of $\lrnd\bW+\IdentityMat_l-\bD\rrnd$, that do not vary from iteration to iteration. 
By linear algebra properties, it is possible to prove that the determinant of $Q(\alpha,\bW)$ can be obtained as:
\begin{equation*}
    |Q(\alpha,\bW)|\,\propto\,\prod_{i=1}^l(1-\alpha\lambda_i).
\end{equation*}
Therefore, from needing to compute a determinant at each iteration, we go to computing the product of $l$ terms.
A comparison of the computing performances of the naive and sparse implementation over a \textit{purely spatial} data set is reported in the Supplementary Material.
We observe a large reduction in the run time, which grows in magnitude as the data size increases. In particular, for more than $300$ locations the naive Leroux takes $\approx 1$ hour, while its sparse implementation requires less than $1$ minute.

The efficient implementation of the \textit{proper CAR} in STAN is available in \cite{joseph2016exact} and has been used in \cite{mingione2022spatio}. The modification of the algorithm to fit the Leroux specification, used in this paper, is instead available on github at \url{https://github.com/PAlaimo/PoiAR_SparseLeroux}.

In particular, we implemented a non-centered parametrization of the space-time random effects specification of Equation \ref{eq:LerouxAR}. Indeed, the NUTS implemented in STAN samples and adapts its tuning parameters in the unconstrained parameters space. Such sampling is more efficient if the posterior is more mildly uncorrelated and approximately Gaussian \citep{betancourt2015hamiltonian}. The non-centered parameterization reduces un-necessary correlation, while discharging some additional (often negligible) burden on the posterior computation. 
In the context of our implementation, this simply encompasses splitting the specification of  the space-time random effects in the following way:
\begin{equation*}
    \begin{aligned}
    &\texttt{Model block}\\ 
    &\bphi^*_t\, \sim \, \Norm\lrnd\bzero\,,\,\bQ\lrnd\alpha, \bW\rrnd^{-1}\rrnd, \quad t=1,\dots,T\\
    &\texttt{Transformed parameters block}\\ &\bphi_1\, = \, \bphi_1^{*}\\
    &\bphi_t\, = \, \rho\cdot\bphi_{t-1} + \sigma\cdot\bphi_t^{*}, \quad t=2,\dots, T
    \end{aligned}
\end{equation*}
Finally, we include a soft sum-to-zero constraint on the random effect in order to keep the intercept in $\bX$ and $\bV$ while maintaining identifiability.
This amounts to adding the following statement in STAN:
\begin{equation*}
    \sum_{\ell ,t}\phi_{\ell t}, \sum_{\ell ,t}\psi_{\ell t}\,\sim\,\Norm(0,\,0.001\cdot n^2),
\end{equation*}
where $n=L\times T$ is the overall size of the random effects\footnote{This is equivalent to having $\bar{\phi}=\frac{\sum_{\ell t}\phi_{\ell t}}{n}\sim \Norm(0,\, 0.001)$}.

Code and examples for the sparse and efficient implementation of the Leroux AR, together with examples of implementation of our Poisson AR model, are publicly available on GitHub at \url{https://github.com/PAlaimo/PoiAR_SparseLeroux}.

\section{Application}
\label{sec:app}
Let $y_{\ell t}$ be the series of weekly cases observed over regions $\ell\in\calS$ at times $t=1,\dots,T$. We wish to model them as discussed in Section \ref{sec:model} and then compare their predictive performance with that of other  models. We consider five possible model specifications.
\begin{enumerate}
    \item[\textbf{a)}] No space-time random effects, i.e.
    $$\phi_{\ell t}, \psi_{\ell t}=0\quad \forall\, \ell, t \text{ in Equations \eqref{eq:epGR} and \eqref{eq:bGR}.}$$ 
    \item[\textbf{b)}] Space-time random effects only on the $\tilde{r}_{\ell t}$, i.e.: 
    $$\psi_{\ell t}=0\quad \forall\, \ell, t \text{ in Equation \eqref{eq:bGR}.} $$
    \item[\textbf{c)}] Space-time random effects only on the baseline $b_{\ell t}$, i.e.: 
    $$\phi_{\ell t}=0\quad \forall\, \ell, t \text{ in Equation \eqref{eq:epGR}.} $$ 
    \item[\textbf{d)}] Space-time random effects on both and covariates, i.e. the full model.
    \item[\textbf{e)}] Space-time random effects on both the terms but no covariates, i.e.: 
    $$\bbeta, \bolds{\eta}=0 \text{ in Equations \eqref{eq:epGR} and \eqref{eq:bGR}.}$$ 
\end{enumerate}

The primary purpose of this modeling  application is to explain the main determinants of the growth rate of  the epidemic. This growth rate is inherently linked to the auto-regressive coefficients $\tilde{r}_{\ell t}$, that approximate it as soon as the baseline $b_{\ell t}$ becomes negligible. Therefore, we are going to include all the covariates in the matrix $\bX$, while $\bV$ contains only the intercept (see Equations \eqref{eq:epGR} and \eqref{eq:bGR}). We consider the following covariates.
\begin{itemize}
    \item The standardized \textit{log-population density}, \textit{log-jsa} and \textit{log-hprice} enter as spatially varying (but time-constant) covariates in $\bX$.
    \item The design matrix $\bX$ alternatively includes one of these two sets:
    \begin{itemize}
        \item[(i) \textit{IND:}] the \textit{Tier} indicator variables;
        \item[(ii) \textit{GMI:}] the standardized \textit{Google Community Mobility Indicators} (GMI) of \textit{Retail and Recreation}, \textit{Residential}, \textit{Workplaces}, and \textit{Grocery and Pharmacy}. Note that the raw index of each district is computed in comparison to a district-specific baseline. Therefore, we perform an inner standardization within each district, i.e. the value of each index is centered and scaled with respect to the mean and standard deviation of the district itself.
    \end{itemize}
    \item The population size divided by $10,000$ is taken as the offset $\text{Off}_\ell,\,\ell\in\calS$, in the baseline rate. The intercept $\eta_0$ indicates the number of hidden cases each $10,000$ people.
\end{itemize}
We note that only the model specifications \textbf{a)} to \textbf{d)} are allowed to have either  one of  the two alternative sets of covariates (i) \textit{IND} or (ii) \textit{GMI} on the auto-regressive coefficients. We exclude having both sets in the same model to avoid multi-collinearity as they would bring redundant information.

The parameters are ascribed typical priors whose variability suite the natural scale of their effects in all settings. Notice that all parameters affect the Poisson rate on the log-scale and therefore the variability is magnified exponentially. We center the intercept $\beta_0$ at a value less than $0$ to favor the identification of a stationary process (on average), that can only temporarily deviate from such stationarity (as an effect of covariates or random effects).
In particular, we assume:
\begin{equation*}
    \begin{aligned}
        &\beta_0\sim\Norm\lrnd-0.5, 1\rrnd,\quad \bbeta_{-0}\sim\Norm_{k}\lrnd\bzero,\, 1\times\IdentityMat_{k}\rrnd,\quad \bolds{\eta}\sim\Norm_\nu\lrnd\bzero,\, 1\times\IdentityMat_{\nu+1}\rrnd\\        
        &\alpha_{\phi}, \alpha_{\psi}, \rho_{\phi}, \rho_{\psi}\sim Unif(0,\, 1),\quad \sigma_{\phi}\sim \Norm^+(0,\, 0.1), \quad \sigma_{\psi}\sim \Norm^+(0,\, 0.1)\\
        &\bw\sim Dir_{\tau}(\bolds{1}_\tau),
    \end{aligned}
\end{equation*}
where $\Norm^+$ denotes the truncated Normal distribution in $(0, +\infty)$, $Dir_{\tau}$ denotes the Dirichlet distribution on the $\tau$-dimensional simplex, and $\bolds{1}_\tau$ is the unit vector of size $\tau$.
The variance of the $\phi_{\ell t}$ random effects impacting on $\tilde{r}_{\ell t}$ is purposefully shrinked more toward zero, as their range of variation shall be smaller than that of the $\psi_{\ell t}$.






\subsection{Simulation study}
\label{subsec:sim}
We consider the \textit{IND} covariate specification and use the observed set of covariates as fixed in all simulations. 
The first design matrix $\bX$ includes $k=9$ terms: the \textit{intercept}, \textit{log - population density}, \textit{log - jsa}, \textit{log - hprice}, \textit{Tier II}, \textit{Tier III}, \textit{Tier IV}, the \textit{Summer effect}, and the \textit{Christmas effect}. The last two artificial covariates have been included to reflect the observed decreased (increased) rate during the summer (Christmas) time. The first spans from the 15th of July to the 31st of August in 2020, the second from the 17th of December 2020 to the 6th of January 2021.
The design matrix $\bV$ instead contains the \textit{Intercept} only.

\begin{figure}[tbp]
    \centering
    \includegraphics[width=0.9\textwidth]{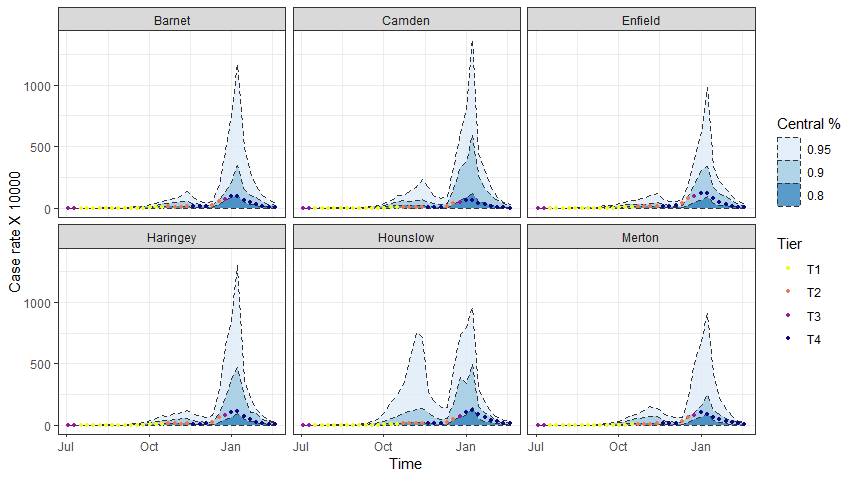}
    \caption{Observed values (dots) and area covered by the simulated trajectories (shaded areas) for six randomly selected districts in the London region.}
    \label{fig:simCov}
\end{figure}

We simulate from the most complex model, i.e. model \textbf{d)}, for $B=400$ times.
The parameters have been chosen so that the simulated trajectories would mimic the observed pattern and comply with our prior setting.
Notice that some coefficients are constant and some others are random, with the latter ones having different values in different simulation data sets.
\begin{itemize}
    \item[$\bbeta^*$] The \textit{Intercept} $\beta_0^*$ has been simulated from  $\Norm(-0.5, 1)$; the \textit{log - population density}, the \textit{log - jsa} and the \textit{log - hprice} coefficients $\beta_1^*, \beta_2^*, \beta_3^*$ are generated independently from the $\Norm(0, 0.1)$ distribution; the remaining coefficients have been set equal to benchmark values that could reflect their natural behavior, i.e. $\beta_3^*=\log(5/6)$, $\beta_4^*=\log(2/3)$, $\beta_5^*=\log(1/2)$, $\beta_6^*=\log(5/2)$, $\beta_7^*=\log(2/5)$, respectively. 
 \item[$\bolds{\eta}^*$] The \textit{Intercept} $\eta_0^*$ has been simulated from $\Norm(0, 1)$.
 \item[$\bolds{\theta}^*$] The parameters governing the two sets of space-time random effects have been generated independently from the following distributions: $\alpha_{\phi}, \alpha_{\psi}\sim Unif(0, 1)$, $\rho_{\phi}, \rho_{\psi}\sim Unif(0, 1)$, $\sigma_{\phi}\sim \Norm^+(0, 0.1), \sigma_{\psi}\sim \Norm^+(0, 0.5)$.
 \item[\textbf{Lags}] Throughout the simulation study we consider the effect of three lags, i.e. $\tau=3$. The weights have been set to $\bw=(0.7, 0.2, 0.1)$ and kept constant in all simulated sets.
\end{itemize}
In the spirit of the prior predictive check \citep{gabry2017visualization}, Figure \ref{fig:simCov} shows the area covered by the central $80\%$, $90\%$, and $95\%$ as compared to the observed values. We can see how the most of the points are included withing the $90\%$ bounds, with the exception of some points just before the Christmas time that, by the way, still lie within the $95\%$ bounds.

We take a connected subset of all the LADCUAs and a slightly reduced time window to speed-up the computing time of the simulation study. In particular, we take the $33$ districts that belong to region of \textit{London} and the weeks that go from the $3$rd of July, 2020, to the $21$st of February, 2021.

We estimate model \textbf{d)} separately on each of the simulated sets keeping  $80\%$ randomly chosen observations in-sample and using the remaining $20\%$ for testing. 
We first evaluate the overall performances in terms of the true parameters recovery and the latent process retrieval.
Tables \ref{tab:simPars1} and \ref{tab:simPars2} report the \textit{root Mean Squared Error} ($\bar{\text{rMSE}}$) and the \textit{average coverage} ($\bar{\text{Cov}}_{95}$) of the corresponding $95\%$ credibile intervals for the $\bbeta$ and $\bolds{\eta}$ parameters over the $B$ simulated sets.
Table \ref{tab:simPars2} does the same for the CAR-AR parameters $\btheta$ governing the two space-time latent processes.

\begin{table}[tbp]
    \centering
    \caption{Root mean square error and average coverage of the $95\%$ credible intervals of the base growth rate and the $\bbeta$ parameters over the $B$ simulated sets}
    \begin{tabular}{l|ccccccccccc}
    \toprule
    \textbf{Metric}     & $\beta_0$ & $\beta_1$ & $\beta_2$ & $\beta_3$ & $\beta_4$ & $\beta_5$ & $\beta_6$ & $\beta_7$ & $\beta_8$\\
         \midrule
    $\bar{\text{rMSE}}$     & $0.10$ & $0.03$ & $0.03$ & $0.03$ & $0.09$ & $0.10$ & $0.10$ & $0.10$ & $0.20$\\
   $\bar{\text{Cov}}_{95}$  & $0.91$ & $0.96$ & $0.94$ & $0.94$ & $0.93$ & $0.97$ & $0.95$ & $0.90$ & $0.92$\\
     \bottomrule
    \end{tabular}
    \label{tab:simPars1}
\end{table}

\begin{table}[tbp]
    \centering
    \caption{Root mean square error and average coverage of the $95\%$ credibile intervals of the $\eta$ coefficients and the lag weights $(w_1,w_2,w_3)$ over the $B$ simulated sets}
    \begin{tabular}{l|cccc}
    \toprule
    \textbf{Metric}     & $\eta_0$ & $w_1$ & $w_2$ & $w_3$\\
         \midrule
    $\bar{\text{rMSE}}$     & $0.07$ & $0.06$ & $0.05$ & $0.03$\\
   $\bar{\text{Cov}}_{95}$  & $0.88$ & $0.92$ & $0.94$ & $0.95$\\
     \bottomrule
    \end{tabular}
    \label{tab:simPars2}
\end{table}

\begin{table}[tbp]
    \centering
    \caption{Root mean squared error and average coverage of the $95\%$ credibility interval of each CAR parameter $\theta$ across the $B$ simulated sets}
    \begin{tabular}{l|cccccc}
    \toprule
    \textbf{Metric}    & $\alpha_{\phi}$ & $\alpha_{\psi}$ & $\rho_{\phi}$ & $\rho_{\psi}$ & $\sigma_{\phi}$ & $\sigma_{\psi}$\\
         \midrule
   $\bar{\text{rMSE}}$    & $0.24$ & $0.17$ & $0.22$ & $0.15$ & $0.04$ & $0.05$\\
    $\bar{\text{Cov}}_{95}$    & $0.93$ & $0.92$ & $0.98$ & $0.91$ & $0.94$ & $0.95$\\
         \bottomrule
    \end{tabular}
    \label{tab:simPars3}
\end{table}

The average rMSE and coverage values are in line with their expected behavior. We notice some minor difficulty in  recovering the exact values of the two baseline intercepts $\beta_0$ and $\eta_0$, probably because of potential partial confounding with the space-time random effects. In any case, the achieved coverage is sufficiently close to the nominal one in all cases.




The performances in the retrieval of the latent space-time random effects are evaluated globally according to the same two metrics. We get $\bar{\text{rMSE}}=0.07$ and $\bar{\text{Cov}}_{95}=0.92$ for $\phi_{\ell t}$ and $\bar{\text{rMSE}}=0.23$ and $\bar{\text{Cov}}_{95}=0.94$ for $\psi_{\ell t}$. Note that that $20\%$ of the random effects are in the test set, and hence those are more difficult to estimate. We note  that the larger values of the $\bar{\text{rMSE}}$ on the $\psi_{\ell t}$ are not surprising as their magnitudes are larger on average than that of $\phi_{\ell t}$ (i.e. $\sigma_\psi>\sigma_\phi)$.

Finally, we evaluate the average in-sample and out-of-sample predictive performances. In this case we consider the average \textit{Relative Mean Squared Error} (RMSE) across simulations and, once again, the average coverage.
The resulting metrics are reported in Table \ref{tab:simPreds}.

\begin{table}[tbp]
    \centering
    \caption{Average root MSE and average coverage of the $95\%$ credibility interval for the in-sample and out-of-sample points}
    \begin{tabular}{l|cc}
    \toprule
    \textbf{Metric}     & \textbf{In-Sample} & \textbf{Out-of-Sample}\\
         \midrule
          $\bar{\text{RMSE}}$    & $0.02$ & $0.10$ \\
          $\bar{\text{Cov}}_{95}$    & $0.99$ & $0.96$ \\
         \bottomrule
    \end{tabular}
    \label{tab:simPreds}
\end{table}

The coverage values are in line with the nominal one for both the in- and out-of-sample observations. The RMSE is practically negligible in-sample and is a bit larger in the  out-of-sample cases. However,  it  is still extremely low as compared to the overall variability of the data. This is typical of flexible model specifications based on random effects that are very good interpolators of the observed data. 


\subsection{Real data application and results}
\label{subsec:res}
The outcome of interest $y_{\ell t}$ of our final application is the series of the weekly cases detected in the $313$ English LADCUAs between the $8th$ of May, 2020, and the $8th$ of March, 2021. The four weeks before the starting date have been discarded as outcome, but are used as lagged counts (down to $5$ weeks behind) for the first weekly records.

In some preliminary attempts, not reported here for the sake of brevity, we have fitted the models for different values of the process memory $\tau$, i.e. the size of the weighted average of previous cases (see Equation \eqref{eq:poiAR1}). We considered $\tau=2,\dots, 5$ and the weights $w_k,\, k=1,2,3,4,5$ were estimated to decay with $k$, reaching values of $\approx 0$ for $k>3$. 
Therefore, we decided to restrict our attention to the case where $\tau=3$. 
As for the simulation setting, we choose $80\%$ of the observations randomly for fitting (in-sample) and use the remaining $20\%$ as testing set to verify the final predictive performances. For each model we run three chains of $8,000$ iterations each, applying a thinning of $2$ to reduce the memory burden of the final workspaces.
All model specifications and settings are compared in terms of WAIC and LOO scores \citep{watanabe2013widely, vehtari2017practical}. Note that the lower the WAIC and the larger the LOO, the better the model fit.
Values of these  criteria are reported in Table \ref{tab:modRank} where  the best scores in each column are highlighted in bold.
\begin{table}[tbp]
    \centering
    \caption{WAIC and LOO scores for the model specifications \textbf{a)} to \textbf{e)} in the covariates settings \textit{\textbf{IND}} and \textit{\textbf{GMI}}}
    \begin{tabular}{l|cccc}
    \toprule
    \multirow{2}{*}{\textbf{Model}}    &  \multicolumn{2}{c}{\textit{IND}} &  \multicolumn{2}{c}{\textit{GMI}}\\
       &  \textbf{WAIC} & \textbf{LOO} & \textbf{WAIC} & \textbf{LOO} \\
                         \midrule
    \textbf{a}   &  $312252$ & $-156105$  & $272202$ &  $-136088$\\
    \textbf{b}   &  $85529.2$ & $-45026.6$  & $85528.0$ & $-45040.01$ \\
    \textbf{c}   &  $84800.2$ & $-44459.6$  & $84434.26$ & $-44315.32$ \\
    \textbf{d}   &  $\mathbf{83348.5}$ & $\mathbf{-44013.5}$  & $\mathbf{83347.5}$ & $\mathbf{-44012.7}$ \\
    \textbf{e}   &  $83520.11$ & $-43817.41$  & $83520.11$ & $-43817.41$ \\
    \bottomrule
    \end{tabular}
    \label{tab:modRank}
\end{table}
The WAIC and LOO scores show identical patterns for models with the two sets of alternative  covariates, with the full model \textbf{d)} having the best scores according to both the metrics. 
Generally speaking, both the GMI and IND setting are valuable in terms of the explanatory power they have about what happened during the second and third wave of the COVID-19 pandemic in England, even with different merits.
However, the latter is advantageous since it depends on exogenous covariates, i.e. policies that can be implemented externally. We can learn about the effect of tiering from the past and use that knowledge to project future scenarios.
This sort of projection is not possible in the case of the \textit{GMI} setting because the values of the \textit{Community Mobility Indices} are not available in advance. 
We further discuss the results from these two models in greater details in the following sections.

\subsubsection{Coefficients of the best model when using the setting \textit{IND}}
\label{subsec:resIND}

First, we focus on the estimation of the covariate effects $\bbeta$ and $\bolds{\eta}$.
There is a particular interest in the estimated incremental intercepts of the different levels of  \textit{Tiers}, i.e. local restrictions imposed by different levels of tiering. Indeed, we aim to check if all classes of restrictions were effective and, if so, by how much with respect to the previous level of restrictions as summarised by the tiering category. Note  that the estimates of the coefficients show the average effect of each component aggregated over all spatial units and over the whole time period.

The estimated coefficients $\hat{\bbeta}, \hat{\bolds{\eta}}$ and the corresponding $95\%$ credibility intervals $CI_{95}$ are reported in Table \ref{tab:estModInd}, together with the $\hat{R}$ to check MCMC convergence. 
\begin{table}[tbp]
    \centering
    \caption{Estimated coefficients, $95\%$ credibility intervals, and $\hat{R}$, the convergence indicator due to \cite{gelman1992inference},  of the $\beta$ and $\eta$ coefficients in the covariate setting \textit{IND}.
    To achieve adequate  convergence $\hat{R}$ should be near 1.}  
    \begin{tabular}{c|l|ccc}
    \toprule
    & \textbf{\textit{IND} coefficients}  &  $\hat{\beta}$  & $CI_{95}$ & $\hat{R}$ \\
    \midrule
    \multirow{8}{*}{$\tilde{r}_{\ell t}$} & Intercept       & $-0.55$ & $(-0.63, -0.47)$   & $1.00$ \\
    & Log-popden                                        & $0.07$  & $(0.05, 0.08)$   & $1.00$ \\
    & Log-jsa                                           & $0.02$  & $(0.01, 0.03)$   & $1.00$ \\
    & Log-hprice                                        & $-0.06$ & $(-0.09, -0.04)$   & $1.00$ \\
    & Tier II                                           & $-0.08$ & $(-0.13, -0.03)$ & $1.00$\\
    & Tier III                                          & $-0.22$ & $(-0.29, -0.15)$ & $1.01$\\
    & Tier IV                                           & $-0.23$ & $(-0.30, -0.15)$ & $1.00$\\
    \midrule
    \multirow{1}{*}{$b_{\ell t}$} & Intercept               &  $0.44$ & $(0.36, 0.53)$   & $1.00$ \\
    \bottomrule
    \end{tabular}
    \label{tab:estModInd}
\end{table}
The estimated intercept is $\hat{\beta}_0=-0.55$, which corresponds to an estimated average (spatially and temporally) growth rate of the detected cases $e^{\hat{\beta}_0}\approx  = 0.58< 1$ in \textit{Tier I}. This means that, when Tier I restrictions are in place, each detected case from the current week will be responsible for slightly more than $1/2$ cases in the next three weeks. 
This estimated rate is rather low, but we must keep in mind this is the rate at which the detected cases reproduce, i.e. those cases that are aware of being infected and are quarantined.
\textit{Population density} is estimated to be positively associated with the growth rate, which is
as expected and agrees with the results of \cite{sahu2022bayesianPaper}. The coefficients of \textit{jsa} and \textit{hprice} show how, apparently, richer regions have been characterized by lower growth rates than poorer ones. The \textit{Tier} restrictions are all estimated to be significantly effective. Nevertheless, there seems to be no significant difference between the effects of \textit{Tier III} and \textit{Tier IV}.
In particular, the average auto-regressive coefficient is $<1$ and the other covariates have negative or small impacts that cannot bring it to surpass $1$. Therefore, we estimate the process to be stationary in all districts and deviations from such stationarity can only by temporarily because of the space-time random effects influence.

The estimated  $\hat{\eta}_0=0.44$ implies an average rate of the hidden cases of $e^{\hat{\eta}_0}=1.55$ cases each $10,000$ inhabitants (see the begenning of Section \ref{sec:app}).


The estimates of the parameters of the space-time random effects $\phi_{\ell t}$ and $\psi_{\ell t}$ are reported in Table \ref{tab:estCarIND}. They can give us quantitative insight on the amount of spatial and temporal dependence present in the data, which was qualitatively visible Figures \ref{fig:MapsCases} and \ref{fig:caseratel}.
\begin{table}[tbp]
    \centering
    \caption{Estimated coefficients, $95\%$ credibility intervals, and $\hat{R}$ for the CAR-AR coefficients governing the two sets of random effects in setting \textit{IND}}
    \begin{tabular}{l|ccc|ccc}
    \toprule
    \multirow{2}{*}{\textbf{\textit{IND} CAR-AR}}  &   \multicolumn{3}{c}{$\bphi$} &   \multicolumn{3}{c}{$\bpsi$} \\
    &   $\hat{\theta}$  & $CI_{95}$ & $\hat{R}$ &   $\hat{\theta}$  & $CI_{95}$ & $\hat{R}$ \\
    \midrule
    $\alpha$       & $0.998$  & $(0.997, 0.999)$ & $1.00$ & $0.98$ & $(0.97, 0.99)$ & $1.00$\\
    $\rho$         & $0.65$  & $(0.60, 0.71)$ & $1.02$ & $0.57$ & $(0.53, 0.61)$ & $1.00$\\
    $\sigma$       & $0.18$  & $(0.17, 0.20)$ & $1.00$ & $0.98$ & $(0.92, 1.00)$ & $1.01$\\
    \bottomrule
    \end{tabular}
    \label{tab:estCarIND}
\end{table}
The high value of the $\hat{\alpha}$'s indicates that the model detects a very strong spatial dependence in both components, which is in line with what is . The estimated temporal dependence is captured by the $\hat{\rho}$'s and it is moderate. Apparently, the the growth rate exhibits a slightly stronger temporal connection. 
The values of two standard deviations $\hat{\sigma}$'s are not directly comparable as the two sets of random effects operate on different scales, but are both contained with respect to the magnitude of the case-rates.

The estimated weight of the three previous weeks on reproduction rate of detected cases is reported in Table \ref{tab:estWIND}.
\begin{table}[tbp]
    \centering
    \caption{Point estimate, $95\%$ credibility intervals, and $\hat{R}$ for the weights $\bw$ governing the influence of the previous three weeks in setting \textit{IND}}
    \begin{tabular}{l|ccc}
    \toprule
    \textbf{\textit{IND} Weights}  &   $\hat{\theta}$  & $CI_{95}$ & $\hat{R}$ \\
    \midrule
    $w_1$       & $0.70$  & $(0.67, 0.73)$ & $1.00$ \\
    $w_2$       & $0.19$ & $(0.16, 0.22)$  & $1.00$\\
    $w_3$       & $0.11$ &  $(0.09, 0.15)$ & $1.00$\\
    \bottomrule
    \end{tabular}
    \label{tab:estWIND}
\end{table}
As expected, the value of the weight $\hat{w}_i$ decays with increasing lag $i$. Approximately $70\%$ of the growth is the contribution of the previous week. Out of the remaining contributions, approximately $20\%$ and $10\%$ are due to lags two and three, respectively. 

The convergence assessment is good for all parameters. We report that there are no divergences or maximum treedepth warnings issues from Stan.

\subsubsection{Coefficients of the best model on setting \textit{GMI}}
\label{subsec:resGMI}

We first consider the estimates of the covariate coefficients $\bbeta$ and $\bolds{\eta}$ in the \textit{GMI} setting.
We aim to check what kind of public behavior (in terms of aggregation in some specific areas) has a positive or negative effect on the growth rate of the detected cases. 

The estimated values and the corresponding $95\%$ credibility intervals $CI_{95}$ are reported in Table \ref{tab:estModGMI}, together with the $\hat{R}$ to check  MCMC convergence.
\begin{table}[tbp]
    \centering
    \caption{Estimated coefficients, $95\%$ credibility intervals, and $\hat{R}$ for the $\beta$ and $\eta$  coefficients in the covariate setting \textit{GMI}.}
    \begin{tabular}{c|l|ccc}
    \toprule
     & \textbf{\textit{GMI} coefficients}  &  $\hat{\beta}$  & $CI_{95}$ & $\hat{R}$ \\
    \midrule
    \multirow{7}{*}{$\tilde{r}_{\ell t}$} &  Intercept        & $-0.68$  & $(-0.75, -0.62)$ & $1.02$ \\
    & Log-pop density                                     & $0.07$   & $(0.05, 0.08)$    & $1.00$ \\
    & Log-jsa                                             & $0.02$   & $(0.01, 0.03)$    & $1.00$ \\
    & Log-jsa                                             & $-0.06$  & $(-0.09, -0.04)$    & $1.00$\\
    & Retail and recreation                               & $0.10$   & $(0.03, 0.17)$   & $1.02$  \\
    & Residential                                         & $-0.12$  & $(-0.22, -0.02)$ & $1.01$  \\
    & Workplaces                                          & $0.07$   & $(0.00, 0.14)$   & $1.03$  \\
    & Grocery and Pharmacy                                & $-0.01$  & $(-0.05, 0.02)$   & $1.01$ \\
    \midrule
    \multirow{1}{*}{$b_{\ell t}$} & Intercept                 & $0.44$   & $(0.35, 0.52)$ & $1.01$    \\
    \bottomrule
    \end{tabular}
    \label{tab:estModGMI}
\end{table}
The intercept is estimated to be $\hat{\beta}=-0.68$, which corresponds to an estimated average (spatially and temporally) growth rate of the detected cases $e^{\hat{\beta}_0}\approx 0.50$. This is not directly comparable to the Tier I intercept of the \textit{IND} setting as, in this context, the intercept represents the average reproduction observed when the indices are equal to the averages of each single district (and not Tier I). It is then reasonable that this value is slightly lower than the corresponding estimate in the \textit{IND} setting.

The \textit{Population density}, \textit{jsa} and \textit{hprice} coefficients are estimated to very similar values of the \textit{IND} setting, and yield the same conclusions.
The coefficients of the \textit{GMI} indices coefficients respect the expected pattern. The epidemic growth inversely proportional to the time people spend at home (i.e. \textit{Residential)}. On the contrary, \textit{Retail and recreation} and \textit{Workplaces} have a positive effect, with the first one more prominent than the second. The \textit{Grocery and Pharmacy} index instead does not show a significant effect (net of the others). This may be a sign that going out for necessities (shopping for food or medicine) does not significantly increment the growth rate of the epidemic, and it is consistent with the finding that Tier IV (T4) was not significantly more effective than Tier III (T3). 



The estimates of the parameters of the space-time random effects $\phi_{\ell t}$ and $\psi_{\ell t}$ are reported in Table \ref{tab:estCarGMI}. The spatial and temporal dependence pattern are very similar to those reported in the \textit{IND} setting, with stronger dependence on the growth rate than on the baseline.
\begin{table}[tbp]
    \centering
    \caption{Estimated coefficients, $95\%$ credibility intervals, and $\hat{R}$ for the CAR-AR parameters governing the two sets of space-time random effects in setting \textit{GMI}}
    \begin{tabular}{l|cccccc}
    \toprule
    \multirow{2}{*}{\textbf{CAR-AR \textit{GMI}}}  &   \multicolumn{3}{c}{$\bphi$} &   \multicolumn{3}{c}{$\bpsi$} \\
    &   $\hat{\theta}$  & $CI_{95}$ & $\hat{R}$ &   $\hat{\theta}$  & $CI_{95}$ & $\hat{R}$ \\
    \midrule
    $\alpha$       & $0.998$  & $(0.997, 0.999)$ & $1.02$ & $0.98$ & $(0.975, 0.999)$ & $1.01$\\
    $\rho$       & $0.63$ & $(0.56, 0.68)$ & $1.05$       & $0.58$ & $(0.54, 0.61)$ & $1.02$\\
    $\sigma$       & $0.18$ & $(0.17, 0.20)$ & $1.04$     & $0.98$ & $(0.92, 1.04)$ & $1.01$\\
    \bottomrule
    \end{tabular}
    \label{tab:estCarGMI}
\end{table}
Finally, Table \ref{tab:estWGMI} reports the estimated weights of previous weeks on growth rate of detected. Also here, the results are consistent with those reported in the \textit{IND} setting.

\begin{table}[tbp]
    \centering
    \caption{Point estimate, $95\%$ credibility intervals, and $\hat{R}$ for the weights $\bw$ governing the influence of the previous three weeks in setting \textit{GMI}}
    \begin{tabular}{l|ccc}
    \toprule
    \textbf{Weights \textit{GMI}}  &   $\hat{\theta}$  & $CI_{95}$ & $\hat{R}$ \\
    \midrule
    $w_1$       & $0.70$  & $(0.67, 0.73)$ & $1.00$ \\
    $w_2$       & $0.19$ & $(0.15, 0.22)$ & $1.00$\\
    $w_3$       & $0.11$ & $(0.09, 0.14)$ & $1.00$\\
    \bottomrule
    \end{tabular}
    \label{tab:estWGMI}
\end{table}

This good degree of agreement between the two alternative settings is re-assuring in terms of the stability of the estimates. We are indeed describing the same phenomenon, even if with a different sets of covariates. 

The convergence assessment is not as good as in the \textit{IND} setting, but is still acceptable since all $\hat{R}\leq 1.05$. This slightly worse quality assessment can be due to  an effect of the strong correlation across the various mobility indices. However, also in this setting there were no warnings regarding  divergences or maximum treedepth from Stan.

\subsubsection{Estimated space-time random effects and prediction performances in the IND setting}
\label{subsec:stre}
We now focus only on the results of the full model with the \textit{IND} covariate setting. 
The analogous discussion on the results of the \textit{GMI} setting are included in the Supplementary Material.

\begin{figure}[tbp]
    \centering
    \includegraphics[width=0.9\linewidth]{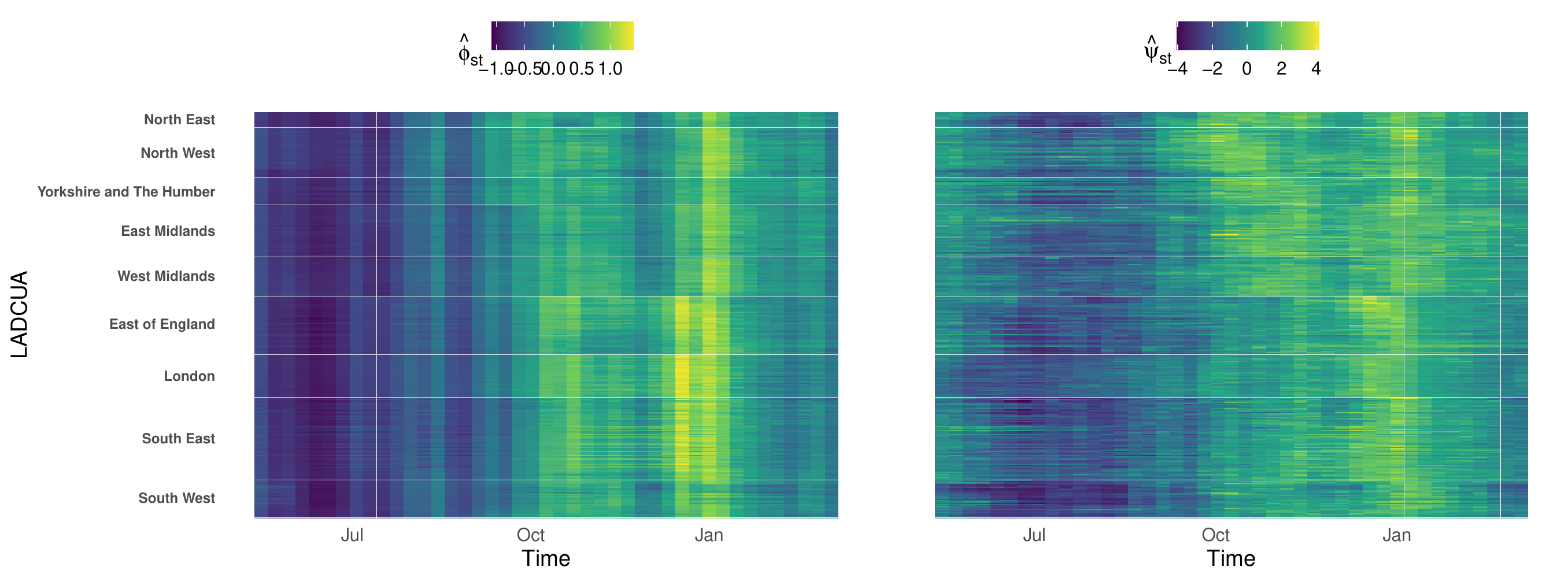}
    \caption{Heatmaps of the space-time random effects $\phi_{\ell t}$ (left) and $\psi_{\ell t}$ (right). All districts are ordered first by region and then within region, from north (top) to south (bottom).}
    \label{fig:streIND}
\end{figure}

The estimated space-time random effects values are reported in the heat-map in Figure \ref{fig:streIND}, where all districts are ordered first by region and then within region, from north to south.

We notice a clear change in the regime of the virus spread between October and December 2020.
It occurred with slightly different timings in different regions, with a clear north-south geographical divide.
According to the model, this occurred both in the reproduction of detected cases and in the rate of the hidden ones, with the former having a very significant spike in the last few weeks of December and first week of January. This last spike can be easily explained through the gathering of families in residential areas during the Christmas period and the high demand for tests that accompanied at that time  as it was recommended by the UK government that people test themselves before mixing with others in Christmas gathering \footnote{\href{https://www.theguardian.com}{https://www.theguardian.com/world/2021/dec/28/covid-test-shortages-threaten-new-years-eve-celebrations-in-england}}.

\begin{figure}[tbp]
    \centering
    \includegraphics[width=0.9\linewidth]{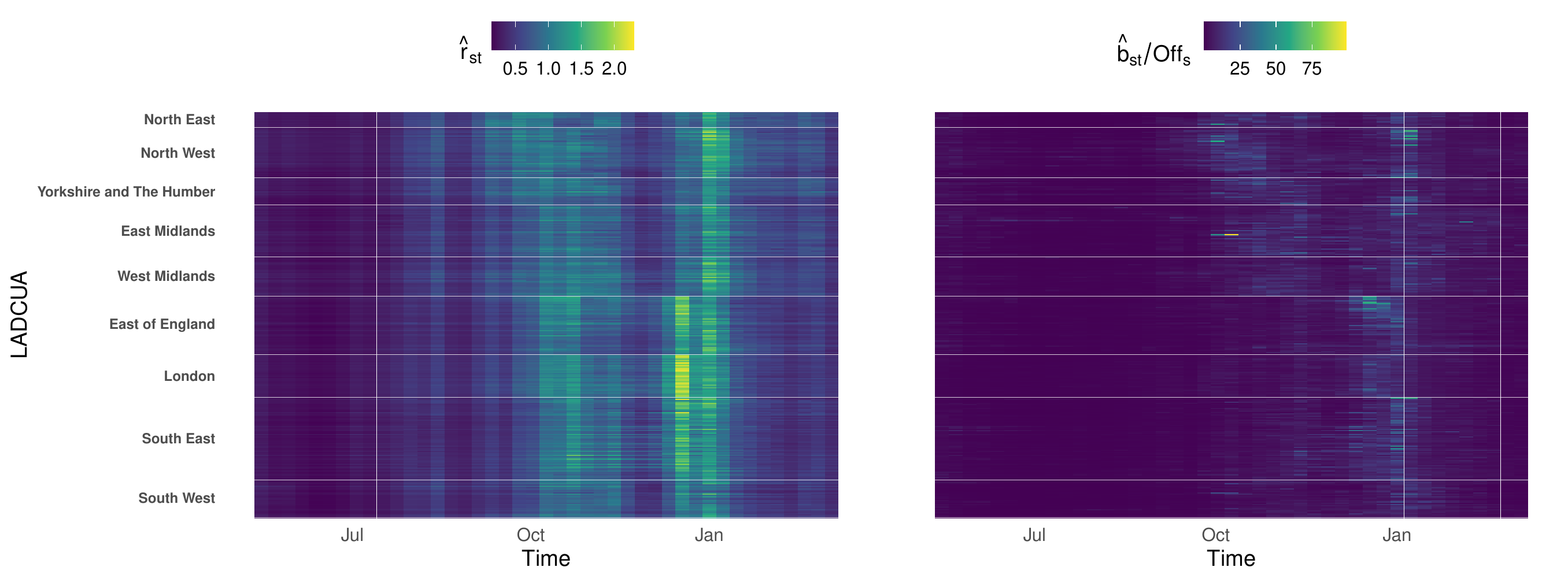}
    
    \includegraphics[width=0.9\linewidth]{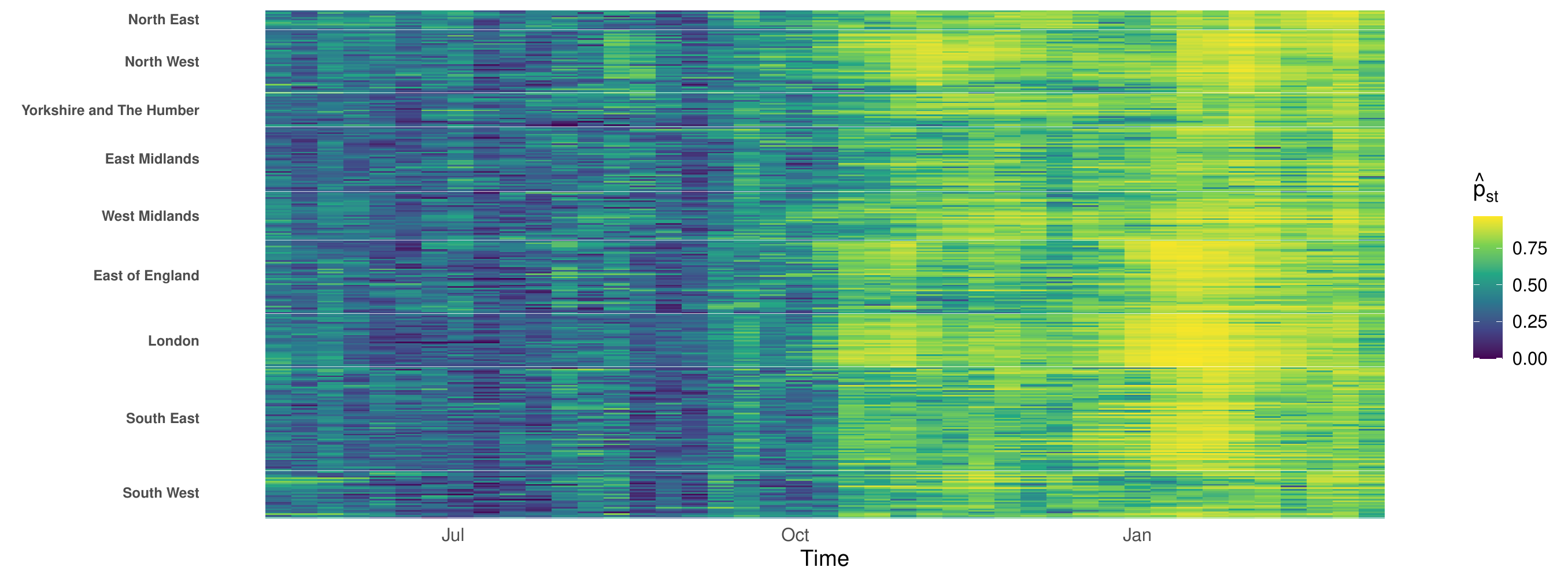}
    \caption{Heatmaps of the estimated reproduction rate $\hat{r}_{\ell t}$ (top-left) and the baseline discounted by the offset $\hat{b}_{\ell t}/Off_s$ (top-right). Heatmap of the estimated proportion of cases due to the reproduction of detected cases and the reproduction of hidden cases $\hat{p}_{\ell t}$.
    All districts are ordered first by region and then within region, from north (top) to south (bottom).}
    \label{fig:ModCompsIND}
\end{figure}

The random effects themselves do not give a clear idea about the relative magnitude of the two components. In order to investigate further, we plot the estimated reproduction rate of detected cases $\hat{r}_{\ell t}$ and the estimated standardized rate of hidden cases $\hat{b}_{\ell t}/Off_{\ell t}$ in Figure \ref{fig:ModCompsIND}. It is also accompanied by the proportion of total cases of each week that could be explained by the first component:
\begin{equation*}
    \hat{p}_{\ell t} = \frac{\sum_{i=1}^{\tau} \lrnd w_i\cdot y_{\ell , t-i}\rrnd \cdot \hat{r}_{\ell ,t}}{\sum_{i=1}^{\tau} \lrnd w_i\cdot y_{\ell , t-i}\rrnd \cdot \hat{r}_{\ell ,t} + \hat{b}_{\ell t}}
\end{equation*}

The third panel of Figure \ref{fig:ModCompsIND} shows a clear shift in the balance between the two components between the first and the second half (before and after October 2020) . Both $\hat{r}_{\ell t}$ and $\hat{b}_{\ell t}$ increase in the second section, but the increase in $\hat{r}_{\ell t}$ is more prominent and its impact on the overall rate $\hat{\lambda}_{\ell t}$ becomes dominant.
This behavior may be explained through two main factors. First, antigen tests became first available in September 2020 and started becoming more and more accessible in the following months. This implied a sensible variation in the testing process, which became easier, faster, and cheaper. This must have increased the size of the detected cases. Second, the diffusion of the \textit{alpha} variant (more contagious and more severe than the original strain) in the UK is estimated to have escalated exactly between October and December 2020 \citep{walker2021tracking}. This has evidently boosted the spreading process and, at the same time, made it more detectable as an effect of the more severe symptoms.

\begin{figure}[tbp]
    \centering
    \includegraphics[width=0.9\linewidth]{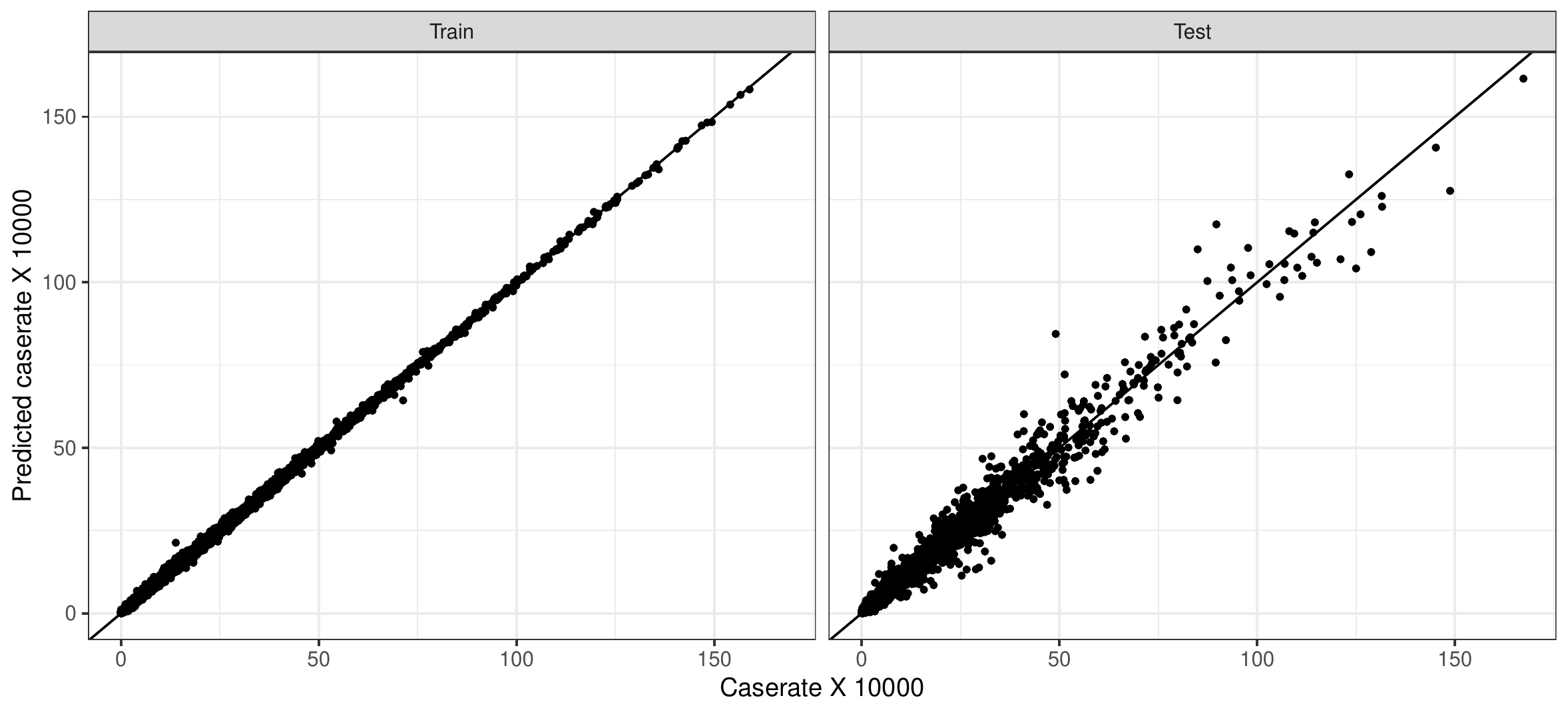}
    \caption{In-sample (left) and out-of-sample (right) predictions of the case-rate per 1000 people (y-axis) and true observed values (x-axis)}
    \label{fig:predsIND}
\end{figure}

We now check the predictive performances in-sample and out-of-sample of model \textbf{d)} under the \textit{GMI} setting.
In-sample and out-of-sample predictions are compared with the true observed values in Figure \ref{fig:predsIND}. 

As expected, the in-sample predictions are almost perfect. It is not surprising for such a flexible model (as already mentioned in the simulation setting). The out-of-sample predictions are also quite accurate, with the points lying around the bisector with comparable variability through the whole domain of the outcome. In particular, the residuals do not show any alarming pattern.

Furthermore, we provide a quantitative assessment by  evaluating the \textit{Relative Mean Squared Error} $\text{RMSE}$  of the point predictions and the average \textit{Coverage} and \textit{Interval Width} of the the $95\%$ predictive intervals $\bar{\text{Cov}}_{95}, \bar{\text{IW}}_{95}$. 
The results are reported in Table \ref{tab:predsIND}.

\begin{table}[tbp]
    \centering
    \caption{\textit{Relative Mean Squared Error} $\text{RMSE}$  of the point predictions and the average \textit{Coverage} and \textit{Interval Width} of the the $95\%$ predictive intervals $\bar{\text{Cov}}_{95}, \bar{\text{IW}}_{95}$}
    \begin{tabular}{c|cc}
    \toprule
    \textbf{Metric}     & \textbf{In-sample} & \textbf{Out-Of-sample}\\
    \midrule
    $\text{RMSE}$            & $\approx 0$ & $0.02$ \\
    $\bar{\text{Cov}}_{95}$  & $0.99$ & $0.94$   \\
    $\bar{\text{IW}}_{95}$   & $64.4$ & $157.5$   \\
    \bottomrule
    \end{tabular}
    \label{tab:predsIND}
\end{table}

\begin{figure}[tbp]
    \centering
    \includegraphics[width=0.9\linewidth]{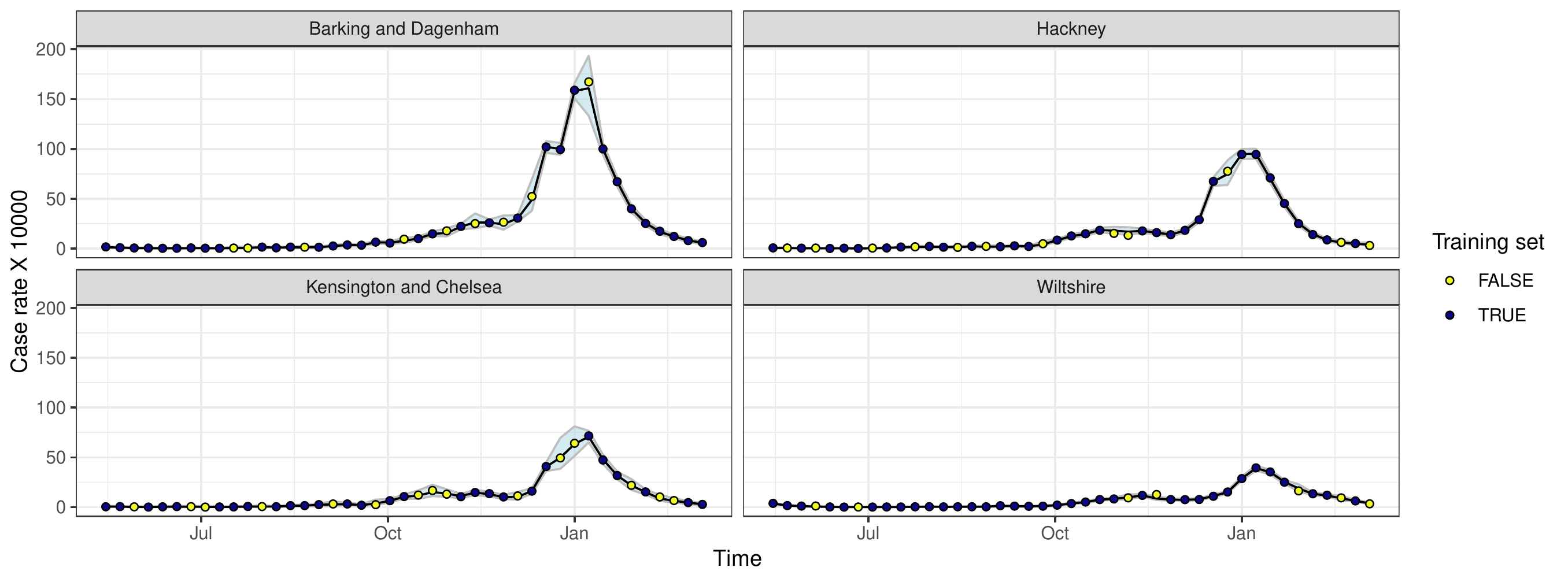}
    \caption{Point-wise predictions and $95\%$ predictive intervals over the observed weekly series of four randomly selected LADCUAS. Points are color-coded for the inclusion in the training or testing set.}
    \label{fig:predsExIND}
\end{figure}

The out-of-sample  $ \text{rMSE}$ is $66.4$, i.e. the average error committed is of $\approx 66$ cases each week per 10,000 individuals.
Figure \ref{fig:predsExIND} shows the prediction performances over the weekly series of four randomly selected districts, with the true data color-coded for inclusion in the train-set.

All the true observations are included within the predictive interval bounds ($100\%$ empirical coverage), with the testing set showing a wider interval because of the inevitable greater uncertainty. 
The point prediction is never far from the true observed value and follows the true trend also in the most complicated out-of-sample cases.

\section{Discussion}
\label{sec:conc}
This paper proposes a space-time extension of the Poisson Auto-regressive model to describe multiple epidemic processes occurring across different areal units. The spatio-temporal variations across the different areas are induced by two different sets of space-time random effects: one controls for the spatio-temporal heterogeneity of the areal epidemic growth rate; the other for the residual process that cannot be explained through the reproduction of the previously detected cases.  The CAR-AR Leroux prior \citep{rushworth2014spatio} is taken as a valid and general choice to describe both the spatial and the temporal dependence in such variations, where the spatiality is induced by the typical neighborhood-based vicinity matrix.
We consider the STAN software package for the estimation of the model. This multi-purpose Bayesian estimation software package implements the No-U-Turn Sampler and is able to fully explore (even high-dimensional) posteriors without requiring much tuning.

Before moving to the real data application, we investigate the identifiability of the model's component through a simulation study. We simulate a pool of datasets with varying parameters according to the proposed model and run the estimation procedure on each of them separately.
The procedure is able to recover the true parameters and quantify their uncertainty, granting coverage of the credibility intervals reasonably close to the nominal level.
In particular, the model is able to detect and account for the various degrees of spatial and spatio-temporal heterogeneity as it is induced by different values of the spatial and temporal smoothing parameters.
The recovery of the two latent processes is also satisfactory. This shows how the model is indeed able to disentangle the auto-regressive from the baseline components, even when there is substantial unobserved heterogeneity in both of them.

The results of the simulation study show that our model is a valid candidate to describe epidemic processes across multiple areas. We apply it to the COVID-19 cases data recorded in the $313$ English LADCUAs between the final tail of the first National lockdown and the end of the third National lockdown ($05/01/2020 - 08/03/2022$).
We compare different model specifications, from the least to the most complex one. The data exhibit substantial spatio-temporal heterogeneity, hence the most complex version is better according to all the considered metrics. All the coefficients' estimates converge well and provide many plausible results. In particular, the results allow for a meaningful interpretation of various external factors. We see how both the Tier indicators (T2, T3, T4) and the GMI indices can well-describe variations in the growth rate of the epidemic. Indeed, they provide comparable inferences but allow for slightly different interpretations. We focus on the tier indicators in the main text, where all tiers' restrictions are estimated to be effective (with various degrees of success) in containing the growth. Nonetheless, tier IV is estimated not to be marginally more effective than Tier III.
Results about the GMI indices show that the more time people spend at home, the less the epidemic grows. In particular, going out for leisure (\textit{retail and recreation}) is the most disruptive of the practices.

At the same time, the space-time random effects capture the space-time variability in the data as the model detects substantial spatial correlation and moderate temporal dependence both in the auto-regressive and the baseline components.
We notice a clear temporal gradient, with the second section of the epidemic (after October 2020) generally stronger than the first one.
We notice a clear-cut north-south geographical divide, slightly less evident during the last wave that appears to be more uniform. In particular, there are clear clusters of over than average or below than average infection rates (South-West for the first, North-West and London for the second).
The flexibility of the model allows it to account for clear outliers through the random effects, without biasing the real underlying estimates and taking on their impact in space and time through the random effects correlations.
We also evaluate the predictive performances of the model, which are good both in-sample and out-of-sample. This shows good coverage but our method does not suffer from unnecessarily high uncertainty. Such coverage is key to addressing public policy as good intervals give way to the construction of best-case and worst-case scenarios.

This work can be extended in multiple directions.
First, it would be interesting to introduce a correlation between the two (now independent) sets of random effects.
Second, the choice of the best size of $\tau$ (i.e. the memory of the growth rate) in this work has been based on heuristics. It could be improved by implementing more robust model selection schemes or direct estimation.
This topic also concerns the choice of what covariates to include in the growth rate and baseline predictors. At the moment, it is demanded to the researcher's sensibility and the research question.

In terms of the input data, we could consider other indices than the tier indicator variables or the GMI. For instance, there is now much enthusiasm about the Covid-19 \textit{stringency} index originally developed by the University of Oxford \citep{hale2020variation}. It is a composite indicator describing the overall level of restrictions in different areas. It is sometimes available also at fine spatial scales, as in the case of Switzerland \citep{pleninger2022covid}.

\section*{Acknowledgments}
This work has been partially supported by \textit{Fondo integrativo speciale per la ricerca} (FISR), grant number FISR2020IP\_00156, and  PON \textit{``Ricerca e Innovazione''} 2014-2020 (PON R\&I FSE-REACT EU), \textit{Azione IV.6 ``Contratti di ricerca su tematiche Green''}, grant number 60-G-34690-1

\bibliography{references}  
\bibliographystyle{unsrtnat}

\end{document}